\documentclass[letterpaper,hyper]{JHEP3}
\usepackage{graphicx}
\usepackage{slashed}
\usepackage{amsmath}
\usepackage{amsfonts}
\usepackage{amssymb}
\usepackage{cite}

\newcommand{\be}{\begin{equation}}
\newcommand{\ee}{\end{equation}}
\newcommand{\bea}{\begin{eqnarray}}
\newcommand{\eea}{\end{eqnarray}}
\newcommand{\eq}[2]{\begin{equation}\label{#1} #2 \end{equation}} 

\newcommand{\alp}{\alpha'}
\newcommand{\Del}{\nabla}
\newcommand{\del}{\partial}
\newcommand{\comment}[1]{}

\renewcommand{\t}[1]{\tilde{#1}}

\renewcommand{\b}[1]{\bar{#1}}
\newcommand{\dgs}{\delta\Gamma^\star}
\newcommand{\ads}{\mathnormal{AdS}_5}
\renewcommand{\L}{\mathcal{L}}
\newcommand{\tev}{\textnormal{TeV}}
\newcommand{\gev}{\textnormal{GeV}}
\title{Warped Kaluza-Klein Dark Matter}
\author{Andrew R. Frey, Rebecca J. Danos,  James M. Cline\\ 
Department of Physics, McGill University, 
Montr\'eal, QC, H3A 2T8, Canada\\
\email{frey,rjdanos,jcline @physics.mcgill.ca}}

\abstract{
Warped compactifications of type IIB string theory contain natural dark
matter candidates: Kaluza-Klein modes along approximate isometry directions
of long warped throats.  These isometries are broken by the full
compactification, including moduli stabilization; we present a thorough
survey of Kaluza-Klein mode decay rates into light supergravity modes and
Standard Model particles.  We find that these dark matter candidates
typically have lifetimes longer than the age of the universe.  Interestingly,
some choices for embedding the Standard Model in the compactification
lead to decay rates large enough to be observed, so this dark matter sector
may provide constraints on the parameter space of the compactification.
}

\preprint{}
\keywords{Cosmology of Theories beyond the SM, Strings and branes phenomenology,
Flux compactifications}

\begin{document}

\section{Introduction}

The search for connections between string theory and observation is
a  longstanding one since such a link would provide a large hint
about the  ultraviolet completion of our effective field theory, the
ultimate  description of physics beyond the Standard Model (SM).  
Recent and ongoing attempts include constructing string models of
inflation (see \cite{arXiv:0901.0265}  for a recent review and more
references) or SM extensions (see \cite{arXiv:0801.0478} for one
example of the Minimal Supersymmetric SM and
\cite{arXiv:0802.3391} for an example of a Grand Unified
Theory).  In this paper, we revisit the argument that dark matter
(DM) may have an intrinsically extra dimensional nature in many
compactifications of string theory.  We will show that the
best-studied compactifications  contain a naturally long-lived DM
candidate particle.  Furthermore, we argue that the decay of the DM
candidate into SM particles can constrain some embeddings of the SM
within the compactification.

Our candidate DM particle is a Kaluza-Klein (KK) mode of the
compactification. KK modes have been considered as DM candidates in
many contexts
\cite{hep-ph/0206071,hep-ph/0207125,hep-ph/0406026,hep-ph/0411053,
hep-ph/0509118,hep-ph/0509119,hep-ph/0701197,arXiv:0805.4210,
arXiv:0811.4356,Park:2009cs,arXiv:0902.0593,Chen:2009gz,arXiv:0907.4993,
arXiv:0908.0899} (see
\cite{hep-ph/0302041,hep-ph/0307062} for an alternative stringy
origin of DM). We focus on a specific corner of the landscape of 4D
solutions of string  theory, warped flux compactifications of type
IIB string theory on  Calabi-Yau manifolds.  These compactifications
contain warped Klebanov-Strassler throat  regions \cite{KS} 
which generate mass hierarchies
\textit{\`a la} Randall-Sundrum \cite{RS}. Since the throats are 
six-dimensional, they can have approximate isometries along the five
angular directions, the Einstein-Sasaki manifold $T^{1,1}$, 
with a corresponding charge that is almost
conserved.  Charged KK modes can therefore be long-lived
cosmologically, as was pointed out by \cite{ky}. Subsequently,
\cite{chentye,dkp} argued that such states at $\tev$  mass, like
thermal WIMPs, can naturally have the correct relic density for DM
candidates, although they only considered KK modes of the graviton,
rather than the whole supergravity spectrum.

Ref.\ \cite{aaron} was the first to try to identify a specific
angular KK DM candidate in the Klebanov-Strassler throat geometry, in
terms of the 10D supergravity spectrum of states and its dimensional
reduction to 5D \cite{italian}.   In this paper, we conduct a more
exhaustive study of decay rates for the  lightest charged bosonic KK modes,
making order of magnitude estimates, extending the work of
\cite{aaron} and improving it using a better understanding of
warped compactifications.  We demonstrate that decays to light 
supergravity fields, including gravitons and light stabilized 
moduli, are all slow compared to the age of the universe for $\tev$
scale DM.   We also consider decays to SM particles in different
brane models for the  SM, a Randall-Sundrum--like D3-brane model and
a D7-brane model.  Finally, we discuss decays involving tunneling to
different regions of the  compactification, along the lines of ref.\
\cite{arXiv:0801.4015}.  

One of our most striking observations is that D3-brane SM sectors can
naturally yield  DM decay rates in an observable range.   Assuming
that the relic density estimates of \cite{chentye,dkp} apply, DM
decay therefore gives direct constraints on the parameter space of
this region of the string landscape.  Since finding such constraints
is a rare opportunity, these models deserve further detailed study. 
We also find that D7-brane models can yield observable
decay  rates, although this statement is more model-dependent than
in the D3 case.   We
emphasize that we did not specifically seek to build models with 
observably large decay rates (in contrast to \cite{arXiv:0902.0008}),
but rather we searched for natural  consequences of common embeddings
of the SM in this class of string compactification.

The plan of the paper is as follows.  In the next section, we give a 
brief review of the class of compactifications we study, including
the warped throat regions of interest.  In section \ref{s:spectrum}
we review and synthesize the spectrum of these compactifications,
including KK modes, light supergravity modes, and SM particles on
D-branes.   Section \ref{s:interactions} discusses the interactions
of our DM candidate with intermediate states and all the most likely
decay products.  We use these interactions to estimate the partial
decay rates in section \ref{s:decay}.  In section \ref{s:discuss}, 
we summarize our
results, compare them to the previous literature on KK modes as DM
candidates in string compactifications, and list directions of
interest for future research.

\section{Review of compactification and warped throat}

The best-understood warped compactifications of string theory are conformally
Calabi-Yau (CY), as described by \cite{BB,DRS,GKP} (see \cite{thesis} for a 
review).  In particular, the 10D metric takes the form 
\eq{wcy}{ds^2 = e^{2A}\eta_{\mu\nu}dx^\mu dx^\nu+e^{-2A} d\t s^2}
for CY metric $d\t s^2$ and warp factor $A$ that depends on the internal 
space.\footnote{For convenience, $\mu$ indices are denoted as raised
with the Minkowski metric and warp factors are counted explicitly.}
These compactifications develop ``throat'' regions where the warp 
factor becomes small and which can act as approximate Randall-Sundrum 
two-brane compactifications \cite{RS,verlinde,cpv,GKP}.  The warp factor is
sourced by 3-form field strengths of the 10D supergravity; these fluxes also
stabilize a subset of the moduli.  The remaining moduli can be stabilized 
either by nonperturbative physics or higher-derivative operators
\cite{kklt,alphap1,alphap2}.  After complete moduli stabilization, there 
should be a small positive cosmological constant and deformations of the
compact manifold, but we ignore these for the most part.

The simplest warped throats in these compactifications were first described
by Klebanov and Strassler (KS) \cite{KS} in the context of gauge/string theory
dualities. KS throats are well-approximated by a warped conifold 
geometry \cite{KW,KT}
away from both their tip and the region that they join to a compactification.
In this regime, the warp factor is such that the Minkowski dimensions and 
radial direction of the conifold form anti-de Sitter spacetime; 
the 10D spacetime factorizes into 
$\mathnormal{AdS}_5\times T^{1,1}$, where $T^{1,1}$ is the 5D base of the
conifold (and is topologically $S^2\times S^3$).  It is important to note that
this approximate geometry is dual to a supersymmetric conformal field 
theory as described in \cite{KW}.

We work in coordinates such that the $\mathnormal{AdS}_5$ metric is
\eq{ads5}{ds^2 = e^{-2kz}\eta_{\mu\nu}dx^\mu dx^\nu +dz^2\ . } With
this radial coordinate, the CY metric in the throat takes the form 
\eq{gtilde}{d\t s^2 = e^{-2kz} \left[dz^2 +\frac{1}{k^2}d\hat
s^2\right]\ ,} where $d\hat s^2$ is the metric of $T^{1,1}$. The
throat runs from $z=0$, where it glues to the compact bulk, to the
tip at $z_0$, and the warp factor at the tip is $w\equiv e^{-kz_0}$. 
At the tip of the throat, the KS geometry is $\mathbb{R}^{1,3}\times
\mathbb{R}^3\times S^3$ with a smooth transition to
$\mathnormal{AdS}_5\times T^{1,1}$.  For the most part, we can simply
replace the smooth tip region by cutting off the geometry at $z_0$. 
In the UV, at $z\approx 0$, the warp factor approaches a constant and
$d\t s^2$ approaches the bulk CY metric smoothly.  As we review
below, the KK scale is $wk$; since we are interested in KK dark
matter, we generally take $wk\sim \tev$.

The $T^{1,1}$ factor has an $SU(2)\times SU(2)/U(1)$ isometry group,
which is preserved by the full KS solution in the noncompact
limit.  This isometry and the associated mode decomposition on the $T^{1,1}$ 
have been discussed in \cite{italian}.  States are labeled by each of their
two $SU(2)$ total spins $j$ and $l$ and by $r=(j_3-l_3)/2$; due to the $U(1)$
quotient, states with the same $j,l$ but different $r$ are nondegenerate.
We  discuss the charges and masses of KK states and other fluctuations 
in \ref{s:spectrum} below. 
The isometry must be broken when the throat is glued to a compact
Calabi-Yau manifold (as those have no isometries), as we describe in
\ref{s:breaking}.  

A note about our conventions:  throughout, we work in the 4D 
Einstein frame appropriate for a stabilized compactification volume.  In 
this frame, the 4D and 10D gravitational scales are related by
\eq{vwarped}{M_p^2\approx M_s^8 V_w\ , \ \ V_w \equiv \int d^6y\sqrt{\t g}
e^{-4A}\ ,}
where $A$ is the background warp factor and the approximation symbol 
indicates that we have redefined the string scale by appropriate factors of
$2\pi$ and the (nearly unity) string coupling.  The relation (\ref{vwarped}) 
is derived from the usual Einstein-Hilbert term for the unwarped 4D metric;
the determinant of the 10D metric contains a factor $e^{-2A}$ when written
in terms of the spacetime metric and $\tilde g$, and there is an additional
factor of $e^{-2A}$ in the contraction of the 4D Ricci tensor.
Note that masses are not
rescaled in transforming from the 10D frame to this Einstein frame.
Also note that the warped volume 
$V_w$ includes the bulk of the CY and is not necessarily 
dominated by the volume of the KS throat; further, we can choose coordinates
such that $A\sim 0$ and $\tilde g_{mn}\sim 1$ (for coordinates of 
dimension length) in the bulk.

\section{Spectrum of fluctuations}\label{s:spectrum} We can now
consider the mass spectrum of excitations of the warped 
compactification, focusing on the behavior in the throat region.  We
are particularly interested in the approximately conserved charge
of each mode.  While most of this section  reviews a range of
results from a  diverse literature, we believe that the synthesis and
interpretation is novel.

We start with a discussion of KK modes in the KS throat, continue with
light supergravity modes (that is,
fields with mass less than the warped KK scale $wk$), 
and finally discuss fields on branes (which represent SM degrees of freedom). 
A summary of the various states we consider is presented in table
\ref{tab:harmonics}.

\TABLE{\centering
\begin{tabular}{|c|c|c|c|}\hline
\multicolumn{2}{|c|}{Particle}&definition & throat wave function\\
\hline
$\gamma\ (\delta\Gamma)$ & $T^{1,1}$ breathing mode &
        eq.\ (\ref{gammaflucts}) 
& ${k^3}M_s^{-4}\, w^{1+\nu}\, e^{(2+\nu)kz}$\\
\hline
$\gamma^\star\ (\delta\Gamma^\star)$ & angular excitation of $\gamma$ &
        eq.\ (\ref{gammaflucts}) & $
{k^3}M_s^{-4}\, w^{1+\nu^\star}\, e^{(2+\nu^\star)kz} \, Y(\theta_i)$\\
\hline
$c$     & universal volume modulus & eq.\ (\ref{univvol}) & constant\\
\hline
$a$       & universal axion & eq.\ (\ref{axionwave}) & $k^{-3}
M_p^{-1}e^{-4kz}$ \\
\hline
$a^\star$ & angular KK axion & \S\ref{s:moduli}& $k^{-3}
M_p^{-1}$[$e^{-4kz}$ or
constant]
$\times Y(\theta_i)$\\
\hline
$\phi, \Theta_3$ & D3-brane fluctuation& 
eq.\ (\ref{D3kinetic3}) & NA \vspace{-3pt}\\
& + superpartner & eq.\ (\ref{D3fermkinetic})& \\
\hline
$\chi, \Theta_7$ & D7-brane fluctuation & eq.\ (\ref{d7kinetic}) & 
constant\vspace{-3pt}\\
& + superpartner & \S\ref{s:d7spectrum},\ref{s:d7coupling}& \\
\hline\hline
\multicolumn{2}{|c|}{Background deformation}&definition & 
throat wave function\\
\hline
$\Delta\Gamma$ & $T^{1,1}$ breathing mode &  eq.\ (\ref{def1}) 
&$w^4\, e^{2kz}\, Y(\theta_i)$\\
\hline
\end{tabular}
\caption[]{The relevant fluctuations of light mass particles and of the
warped background.
In all cases, $Y(\theta_i)$ stands for
$Y_{(1,0,0)}(\theta_i)$ or $Y_{(0,1,0)}(\theta_i)$   }
\label{tab:harmonics}
}

\subsection{KK modes}\label{s:kkmodes}

In this section, we review the wave functions and spectra of the 
lightest KK modes, particularly looking for the lightest mode with angular
charge.  Assuming a thermal history for the KK modes (as in \cite{chentye}),
angular charge should accumulate in this state (for some caveats, see
\ref{s:moduli} below), so it is our DM candidate.

In general, the lightest KK modes are localized in the longest
KS throat, at least when the warped KK scale is less than the 
curvature scale of the bulk CY (see, for example, 
\cite{hep-th/0006191,firtye,chentye,fm}).\footnote{In any event, delocalized 
KK modes would not preserve
angular charge because the bulk CY does not respect the throat's isometries.}
Therefore, we only need to be concerned with the form of KK modes in the
throat, and we can proceed by dimensionally reducing on $T^{1,1}$ and 
then $\ads$ with boundary conditions at $z=0$ and $z_0$. 
The boundary conditions at $z=z_0$ control the mass, so the 
KK scale for localized modes is set by $wk$, though the 4D masses of the 
KK modes are largely insensitive to the form
of the boundary conditions \cite{aaron}.

We thus need to know the contributions to the 5D mass of various fields.  These
include contributions both from the KK reduction on $T^{1,1}$ and from
the moduli stabilization.  The angular motion on $T^{1,1}$ gives contributions
of order $\Delta (m_5^2)\sim k^2$ (derived in detail in \cite{italian}); 
further, motion on the ``$S^2$ factor'' of $T^{1,1}$ generates a similar 
mass due to a centripetal barrier near the tip of the throat \cite{firtye}.  
Moduli stabilization 
includes a classical component due to 3-form flux, which may be as large
as $\Delta (m_5^2)\sim k^2$ or much smaller, depending on the interaction of
the 10D field with the flux (see \cite{fm} for example).  Nonperturbative or
$\alp$ corrections to the compactification can also generate a mass; we
assume that these are small compared to the $\ads$ scale $k$ in order to assure
that these corrections do not deform the geometry too much.  

Note that, even though the precise KK spectroscopy of the throat is known
\cite{italian}, the identity of the lightest charged KK mode is highly
model-dependent (note that the discussion of \cite{aaron} only considered
geometric KK reduction).  In practice, then, we choose a charged mode
as a proxy for the true lightest charged state.  For the purposes of 
order-of-magnitude estimation, this should be reliable as long as the 
dominant decay mechanism is the same, and we  
elaborate guidelines for converting our decay rates to alternate possibilities.

Fortuitously, the simplest mode to consider is also the mode identified
as the lightest charged state using purely geometric KK reduction.  The
lightest states are the breathing mode of the $T^{1,1}$ (and an associated
fluctuation in the warp factor which we can ignore for order-of-magnitude
calculations) and an axion partner, 
both with $(j,l,r)=(1,0,0)$ or $(0,1,0)$ \cite{aaron}.
We can parameterize the breathing mode by deforming the CY metric $d\t s^2$
as 
\eq{gtdef}{d\t s^2 = e^{-2kz} \left[ dz^2 +\frac{e^{2\Gamma(z,\theta)}}{k^2}
d\hat s^2\right]\ .}
The charged KK modes are small fluctuations of $\Gamma$, which we denote
$\dgs$ (the $\star$ representing angular excitation). These fields
saturate the Breitenlohner-Freedman bound \cite{bf} in $AdS_5$, and the
4D mass is of order $wk$, the warped KK scale.   $\Gamma$ also has KK modes
with no angular motion, which are massless in 5D.  Those fluctuations are
actually part of the universal volume modulus, but they also contain
radial KK excitations denoted $\delta\Gamma$ with 4D mass of approximately 
$wk$ (somewhat larger than the mass of $\dgs$).  These uncharged 
KK modes are important as intermediate states in decay amplitudes of 
$\dgs$.

We also need the properly normalized wave function for the KK modes on 
the compactification in order to write interactions in terms of 4D fields,
which we  denote with lower case $\gamma$.  
Massive KK modes are localized in the KS throat, 
predominantly near the tip.  In fact, the wave functions of the lightest
radial KK modes are given by 
growing exponentials in $z$ to a reasonable approximation \cite{aaron}, 
which are
sufficient for our estimates.  For example, a breathing mode fluctuation
with angular momentum $(j,l,r)$ has an approximate wave function
\bea
\delta\Gamma_{(j,l,r)}(x,z,\theta) &\approx& 
\frac{\gamma_{(j,l,r)}(x)}{M_p}\left(
\frac{V_wk^6}{V_{T^{1,1}}}\right)^{1/2}w^{1+\nu}Y_{(j,l,r)}(\theta)
e^{(2+\nu)kz}\nonumber\\
&\approx& \frac{k^3}{M_s^4}\gamma_{(j,l,r)}(x)
w^{1+\nu}Y_{(j,l,r)}(\theta)e^{(2+\nu)kz}\label{gammaflucts}\eea
within the throat, where $\nu^2=4+m_5^2/k^2\sim 1$ for 5D mass 
$m_5$ and the factor of $M_p$ is needed for canonical normalization of the
4D field 
$\gamma$.\footnote{The true radial wave function is a product of an exponential
and a sum of Bessel functions, but \cite{aaron} showed that the pure
exponential behavior is a good approximation for the lowest modes in the
radial KK tower, at least for order-of-magnitude calculations.}
$V_{T^{1,1}}$ is the angular volume, an order unity constant, which we
 henceforth drop.
Considering only the angular dimensional reduction, we would obtain
 $\nu^\star\equiv
\nu(\dgs)=0$ and $\nu\equiv\nu(\delta\Gamma)=2$; however
these values are increased by flux contributions.  
Given the spacing of geometric
5D masses, it seems unlikely that $\nu^\star$ is much larger than 
$2$ for the true lightest
charged state.  In addition, the wave function dies off exponentially 
quickly outside the throat, so we can treat it as zero in that region.
The normalization is fixed up to order 1 factors by the condition that
\eq{normalization}{\frac{1}{V_w} \int d^6y\sqrt{\t g}\, e^{-4A} \delta
\t g_{mn}\delta\t g^{\widetilde{mn}} =\frac{\gamma^2}{M_p^2}\ ,}
as given in \cite{stud}.  The wave function (\ref{gammaflucts})
of course ignores the fact that some angular modes must vanish at the tip
of the throat \cite{firtye}, but we can account for that effect by hand.

\subsection{Moduli and other light supergravity modes}\label{s:moduli}

The moduli and light fields of the compactification have been a topic
of  some interest in the literature, due to their relevance for the
effective field theory of warped compactifications (see, for example,
\cite{fp,thesis,gm,fm,stud,dt,ftud}). Our main interest in light
fields will be as decay products of our DM candidates (we assume that
massive moduli can decay rapidly to SM particles); they may also
be  obstructions to the cosmological population of the KK DM
candidates, but we leave that possibility for future studies.  In
this discussion we consider light states to be all states with 4D
mass significantly less than the warped $\ads$ scale $wk$. These are
states that are massless before moduli stabilization effects.  As
long as their masses remain light, their wave functions are expected
to be similar to what they would be without moduli stabilization
\cite{fm}; semiclassically, when their mass reaches $\sim wk$, it is
energetically favorable for their wave functions to localize near the
tip of the throat --- but it is not at a lower mass.

To identify the light modes,
we start by asking what states might be massless if we ignore moduli
stabilization effects.  Since the wave functions of light modes typically
spread through the entire compactification, they cannot be reliably studied
just by considering dimensional reduction in the throat in a simple way.
Specifically, from the point of view of the $\ads$
throat, the boundary conditions at $z=0$ mix many different fields, which
leads to surprising behavior.  Nonetheless, the approximate isometry of the
throat ensures that we can break down fields by angular charge sectors 
in the throat.  Furthermore, even though the radial wave functions may not
take the form suggested by pure $\ads$ reduction, all possible cases of light
fields are represented by fields with vanishing 5D mass as given by 
reduction on $T^{1,1}$.  We can proceed by 
examining angular charge sectors one at a time.

The most familiar sector is likely the uncharged one, which has 
wave functions uniform on the $T^{1,1}$.  For instance, all warped CY 
compactifications have an overall volume modulus and an associated axion
(sometimes called the universal volume modulus and universal axion 
respectively).  In addition, a conifold throat has two moduli associated 
with the two ways to resolve the singular conifold point; these are identified
as Betti forms in \cite{italian}.  For the KS throat, the modulus 
associated with the small resolution of the conifold is lifted by topology
(essentially it is incompatible with IR boundary conditions), so it has only
higher KK modes.  The modulus associated with the deformation of the
conifold is stabilized by the 3-form flux on the throat.  While this could
still be a light mode if the compactification is large, we  focus on
the universal volume modulus and axion due to the relative simplicity of 
their profiles in the extra dimensions. There may, in fact, be other
uncharged moduli, depending on the CY manifold, 
but we take the universal volume
modulus and axion to be representative of them as well.

By now, the universal volume modulus is well-understood \cite{gm,ftud}.
However, it is not described by an overall rescaling of the CY 
metric $d\t s^2$ accompanied by an appropriate fluctuation of the warp factor; 
rather, it is most simply described by a shift of the warp factor
\eq{univvol}{e^{-4A}\to e^{-4A}+\frac{c(x)}{M_p}}
along with metric fluctuations that depend on the
spacetime derivatives of $c$, the volume modulus (given here with 
canonical normalization up to factors of order unity).  In particular,
note that $c(x)$ is independent of the compact dimensions.  

Similarly,
the universal axion is a fluctuation of the 4-form potential based on the
K\"ahler form of the CY and can be written with all four legs in the compact
dimensions \cite{ftud}.  
This involves multiple components of the 4-form when written in
a given set of coordinates, and all these components are required to 
produce a single massless 4D field (on a CY with multiple K\"ahler moduli,
some of these components could be independent fields, 
but not if there is only the universal volume modulus).  
These modes are also accompanied by 
fluctuations of other supergravity field strengths.  However, we are only 
making rough order-of-magnitude calculations, and we have checked that 
these effects are unimportant for our purposes.  Ignoring those other 
field strengths, it is a simple calculation to see that
the 4-form scales like $e^{-4kz}$ in the throat and 
like a constant in the bulk \cite{ftud}. Picking a sample component, we
have 
\eq{axionwave}{C_{z\theta\phi\psi}\approx e^{-4kz}\frac{a(x)}{k^3 M_p}\ 
\textnormal{(throat)}\ \textnormal{or}\ \approx \frac{a(x)}{k^3 M_p}\ 
\textnormal{(bulk)}\ ,}
where we define the scale of the angular directions as $k$ following
(\ref{gtilde}) and use canonical normalization.  The kinetic term is a 
6D integral dominated by the bulk, which yields the normalization.

As discussed above,
this modulus is stabilized by nonperturbative or $\alp$ effects, but
its mass should be less than the warped KK scale to ensure that the 
corrections do not deform the geometry too strongly.  For low mass, the
wave function should be essentially the same as the classical wave function
\cite{fm}.  Note, though, that the wave functions are not constants in the
throat, as suggested by modeling the throat as $\ads$ with simple 
boundary conditions.
The reason is that the moduli are combinations of many 10D fields, which 
play off of each other in a complicated manner and are related by more 
complicated boundary conditions in the $\ads$ picture.

We now turn to the $(j,l,r)=(1,0,0)$ and $(0,1,0)$ charge sectors, which
are identical in the $\ads\times T^{1,1}$ geometry.  Just considering the
KK masses from the angular reduction, as discussed in \cite{italian}
there are several massless states. These
states include a 4D vector descending from the RR 4-form potential and scalars
from the metric and 4-form potential taking the form of a K\"ahler modulus.
Again, the $\ads$ dimensional reduction is not entirely reliable, but 
both of these types of massless states are possible on a general CY 
compactification.   Either charged vectors or charged moduli can remove the 
angular charge during cosmological evolution.  These would be the lightest
charged states, so thermal equilibrium would eventually drive all the charge
into the light states.  For exactly massless states, this would remove the
charge completely, as radiation redshifts more quickly than matter.  However,
even stabilized charged moduli present a problem.
As long as the 4D mass of these is below the warped
KK scale $wk$, their wave functions should be spread throughout the bulk of
the compactification.  Since the isometry of the throat is strongly broken
in the bulk, these moduli are allowed to have interactions that 
violate angular charge conservation.

However, the
zero modes of the massless vectors are projected out by O3-planes of string
theory, so we can restrict to compactifications with O3-planes. Alternately,
the particular angular harmonic associated with the massless vector may not
be associated with a massless mode on the bulk CY.  In either case,
the boundary conditions at the UV end of the $AdS_5$ throat are incompatible 
with the zero mode, so only higher radial KK modes are allowed.
Analyzing the cosmological transfer of angular charge
into these vectors is an interesting question for future studies of the 
detailed cosmology.  In this paper, however, we simply assume that
these vectors are projected out by the bulk CY physics (possibly an 
orientifold plane).

The light scalars in this charge sector take the form of K\"ahler moduli of 
the CY manifold, a metric scalar and a 4-form axion.  
It is well-known that the number of K\"ahler moduli
is controlled by topological invariants of the CY; if there is only one 
K\"ahler modulus, these zero modes must be projected out, as discussed for the
vector above.  On the other hand, if the CY has multiple K\"ahler moduli,
it is possible that these states can remain massless at the classical level.
Furthermore, the 3-form flux does not stabilize these moduli.  
These moduli are stabilized either by $\alp$ corrections or 
nonperturbative effects.  Again, if this mass is less than the warped KK
scale $wk$, these moduli could provide a cosmological sink for the charge.  
While we leave a discussion of this possibility for future work,
we estimate the decay rate of charged KK modes into these ``charged'' 
moduli on the chance that their scattering rates are too weak to permit
thermalization.

In order to calculate decay rates, we need the wave functions of these
moduli.  Treating them as 5D fields and dimensionally reducing on $\ads$
indicates that these modes should have constant profiles in the KS throat.
However, it is possible that, like the universal volume modulus and axion,
the true zero modes are a mixture of multiple components, which can change
the radial behavior.  An additional complication is that the appropriate 
tensor angular harmonics on $T^{1,1}$ are unknown.  Therefore, we 
use axions (potential decay products themselves) also as proxies for moduli.
For the axion,
we consider both a constant wave function and one that scales like
$e^{-4kz}$, like the universal axion.  In all cases, the wave functions should
be approximately constant in the bulk (because they are given by harmonic
forms).

There is a final possibility for light scalars.
The $(j,l,r)=(1,1,\pm 2)$ charge sectors also have massless scalars
after reduction in the angular directions.  Assuming that these modes are 
massless after matching to the bulk CY metric, they are pairs of metric 
scalars, as expected for complex structure moduli.  Again, we should not
take the $\ads$ dimensional reduction literally, but we can also consider the
case of charged complex structure moduli.  These moduli
are generically stabilized by 3-form flux, like the deformation modulus of the
conifold.  Like the deformation modulus, these may be heavy or 
light fields, but
we  consider other moduli as proxies for them, since we are just working
to order of magnitude.

We conclude this section by  commenting on effects of moduli stabilization 
on wave functions in the throat.  As mentioned 
earlier, if the mass scale of moduli stabilization is below the warped scale
$wk$, the wave function should not be much distorted from the massless case.
However, at masses more than $wk$, the mode should accumulate near the bottom
of the longed throat, behaving essentially like an excited KK mode \cite{fm}.
Therefore, in those cases, we can simply use our results without including
light supergravity states.  There is one caveat, however.  For moduli
stabilized by quantum mechanical or $\alp$ corrections, such a high mass 
may or may not
indicate that the throat has been deformed strongly by the physics that
stabilize the moduli, so it is possible that the geometry is not a small
perturbation of the KS background, breaking our assumptions 
(specifically, when we discuss deformations of the throat later, we assume 
that those deformations are small).  This may be an important concern for
high-scale inflation models; however, our results should be parametrically
similar as long as the throat remains approximately isometric.

\subsection{Brane modes}

In the context of IIB string theory compactifications, the Standard Model (SM)
is typically understood as being supported on a D-brane or stack of D-branes,
either D3-branes or D7-branes in warped CY compactifications.  We are
interested in the case in which the SM branes are located in the same KS
throat as the KK modes of interest; this is somewhat natural due to the 
$\tev$ scale of the throat.  It is also possible that the branes are located
in the bulk (for D7-branes) or another throat.  In that case, any warp
factor quoted in this subsection or \ref{s:braneints} 
should be replaced by that for the appropriate throat.

\subsubsection{D3-brane Standard Model}
The brane scalars correspond to transverse motion of the D-branes; these are
the SM Higgs fields or scalar superpartners of the SM fermions (either of
which can decay quickly to SM fermions).  
The kinetic terms just follow from the DBI action on the branes; in the case
of D3-branes, it is
\eq{D3kinetic}{\mathcal{L} 
= -\frac{1}{2}\frac{\mu_3}{g_s} \textnormal{tr}\left( 
\t g_{mn} \del_\mu \phi^m \del^\mu \phi^n\right)\ .}
The brane scalars are just the coordinates of the brane on the 
conifold and the trace is over the non-Abelian degrees of freedom on the 
branes. We consider a single angular coordinate (along an equator
of the $S^3$ at the tip) and 
reduce to a single diagonal component of the gauge matrix.
With metric (\ref{gtilde}), we find
\eq{D3kinetic3}{\L \approx  -\frac{1}{2}\frac{\mu_3}{g_s} \frac{w^2}{k^2}
\left(\del\phi\right)^2\ ,}
so the canonically normalized scalars are rescaled by $\sqrt{\mu_3/g_s}(w/k)$.
(Alternately, we can change to Riemann normal coordinates at the brane
position, in which case there is no need to rescale by $w/k$.)

To see whether the brane scalars give a representative decay rate,
we also check for decays into D3-brane fermions.
From \cite{hep-th/0202118} (see also
\cite{hep-th/0311241,hep-th/0312232}), the fermion kinetic term on a D3-brane
at $z_0$ is
\eq{D3fermkinetic}{\L=\mu_3  w^4\left[-\frac{1}{2}
w^{-1}\b\Theta_3\slashed{\del}\Theta_3 \right]\ ,}
where $\mu_3$ is the brane charge and $\Theta_3$ is 
a 10D fermion that contains 
all the 4D fermionic degrees of freedom. In the sequel, we use 
$\Theta_3$ 
to represent one 4D fermion.  Up to factors of order unity, then, the
canonically normalized fermions are $\sqrt{\mu_3}w^{3/2}\Theta_3$.

Finally, there are worldvolume gauge fields (the SM gauge fields) on the
D3-branes.  We do consider decays into the gauge fields, however.

\subsubsection{D7-brane Standard Model}\label{s:d7spectrum}

We now consider the case in which the SM lives on D7-branes extended
in the throat.  Our results are largely insensitive to the embedding
of the D7-brane in the compactification, but we show how the D7-brane
kinetic term (\ref{d7kinetic}) matches that for a specific  embedding
in appendix \ref{s:d7metric}.  We assume  that the D7-branes do not
extend to the bottom of  the throat (that is, the branes extend to a
maximum radial coordinate of  $z_1\ll z_0$, where $w_1=e^{-kz_1}$). 
Furthermore, like our treatment of the throat, we replace the smooth IR
tip of the brane with boundary conditions at $z_1$. Except near the
tip of the brane, $z$ lies along the brane, so the  transverse
coordinates lie in the $T^{1,1}$.

In the case of the D7-brane, there is a single complex scalar field $\chi$,
corresponding to fluctuations around the static embedding.  The kinetic term 
of the brane scalars is controlled by the metric 
pulled back to the fluctuating brane.  
Additionally, the light scalars on a D7-brane have a constant profile
in the compact dimensions of the brane (which agrees with \cite{d7susy} for
D7-branes in a KS throat).  These are the appropriate
final states to consider for decays of the dark matter candidates: excited
KK modes of the D7-brane are more massive than the gravitational KK modes
we consider (D7-brane KK mode masses are set by the warped scale at $z_1$
suppressed by flux quantum numbers, and this suppression is insufficient to 
reduce the mass to the warped KK scale at $z_0$). 

It is true that the light scalar $\chi$ must in fact be stabilized, 
although we  treat it as massless.  First, the mass scale of $\chi$
may be much less than the warped scale at the bottom of the D7-brane $w_1k$;
indeed, if $\chi$ is stabilized by supersymmetry breaking, this situation
may be natural.  In that case, the $\chi$ wave function is 
approximately constant, following the same logic as for moduli.  
Furthermore, we want to estimate the interaction of the
KK modes with SM particles, which are very low mass compared to $w_1k$.  
Therefore, we are justified in treating the $\chi$ particles as massless.

For brane scalars that depend only on the external spacetime 
$\chi=\chi(x^\mu)$,
we can approximate the pulled back metric as 
\eq{d7pullback}{ds_8^2\approx \left(e^{2A}\eta_{\mu\nu} +2e^{-2A}
\t g_{\chi\bar\chi}\del_{(\mu}\chi\del_{\nu)}\bar\chi\right) dx^\mu dx^\nu
+e^{-2A}\t g_{ij} dy^i dy^j\ ,}
where $y^i$ are the four compact coordinates along the brane.  There are
generically off-diagonal terms of the form $\del_\mu\chi dx^\mu dy^i$, 
but these
lead to parametrically similar contributions to the kinetic term (which
we see for a specific example in appendix \ref{s:d7metric}).  We choose 
coordinates in the stabilized compactification such that $\chi$ matches from
the bulk to the throat, and we treat $\t g_{\chi\bar\chi}$ as 
approximately constant and order unity in the bulk.  
Then the approximate kinetic term from the DBI action is
\eq{d7kinetic}{\L_\chi \approx -\frac{\mu_7}{g_s}\int d^4y\sqrt{\t g_4}\,
e^{-4A}\t g_{\chi\bar\chi}|\del\chi|^2\ .}
In the throat, $\tilde g_{mn}\propto e^{-2kz}$, so the integrand dies off
farther into the throat.  That means the kinetic term is dominated by the 
part of the D7-brane that extends into the bulk CY manifold.  Therefore,
\eq{d7kinetic2}{\L_\chi\approx -\frac{\mu_7}{g_s}V_{4w}
|\del\chi|^2\ ,}
where $V_{4w}$ is the warped 4-volume along the brane.  In the bulk, where
the warp factor is $A\sim 0$,  we can approximate it 
by $V_{4w}\sim V_w^{2/3}=(M_p^2/M_s^8)^{2/3}$.
This gives us immediately the canonical normalization for $\chi$.

Similarly to our treatment of the D3-brane Standard Model, we do not
consider the brane gauge fields (which include 4D scalars for a
D7-brane).  We do, however, consider the D7-brane fermions,
estimating their interactions based on those of the brane scalars, as
explained below.

\subsection{Supergravity fermion KK modes}
\label{sfkkm}

The reader may note that we have not discussed modulini or 
KK modes of the supergravity fermions, despite the fact that the lightest
charged state might very well be fermionic.  The reason is two-fold.
First, due to the fact that the simplified background we consider is 
supersymmetric, we expect that the decay rates of fermionic states should
be similar to those of bosonic KK modes.  Second, the calculation of 
fermionic decay rates is technically more difficult, as we now explain.

The difficulty lies in the simple fact that the decay products must always
contain a fermion.  Considering purely supergravity interactions is not
difficult, although some care is needed in decomposing the 10D spinors
into 4D spinors to find off-diagonal mixings and Yukawa couplings.  While
it is not as familiar as the bosonic action, the 10D supergravity 
is known for fermions (see \cite{Howe:1983sra,Schwarz:1983qr} and subsequent
papers).  However, while interactions between brane fermions and bosonic
supergravity fields are known for D3- and D7-branes, mixings and mixed Yukawa
terms between brane and supergravity fermions are not known in generality,
particularly with warping included.  
To find these interactions, it is necessary to expand the 
$\kappa$-symmetric D-brane action to include more terms in the 10D 
superfields, as done in \cite{hep-th/0502059} for gravitino-brane fermion
bilinears.  While this is an
interesting project, it is beyond the scope of the present work, and we
leave it to the future.

\section{Interactions}\label{s:interactions}

In this section, we  describe the interactions of the KK modes,
especially $\dgs$ and $\delta\Gamma$,
which we  take to be representative of general angular KK modes in the
KS throat, particularly the lightest one.  We begin by discussing 
quadratic mixings and three-point interactions in the supergravity action
and then move to interactions on branes.  One or two other relevant 
interaction terms are also presented.

\subsection{Supergravity interactions}
All the interaction terms in
the KS throat must be neutral under the approximate isometry, though branes
that explicitly break the isometries can support charged interaction terms.
As a result, deformations of the KS throat that break the approximate 
isometry, communicating the asymmetry of the bulk CY, is important to
our analysis, so we discuss them first.

\subsubsection{Isometry-breaking deformations}\label{s:breaking}

As we discussed above, compactification effects break the isometry of 
the KS throat, if for no other reason than that compact CY manifolds have no
isometries.  However, we demand that these be small perturbations to the 
throat geometry as the KS throat background would no longer be valid.

Since these deformations have nontrivial profiles on the 
$T^{1,1}$ in order to break the isometry, they are therefore also 
classified by the quantum numbers $(j,l,r)$.  The radial behavior of these 
deformations is governed by the conformal symmetry of the $\ads$
geometry (with small corrections in the full KS geometry), as 
has been discussed by \cite{aab}.  In short, the radial behavior of the 
deformations is exponential in $z$ with a coefficient determined completely
by the angular quantum numbers.  In fact, since the simplified geometry is
dual to a conformal field theory, even $\alp$ and quantum corrections to the
supergravity cannot change the radial behavior of the 
deformations!\footnote{If we include the full KS throat background geometry,
the subleading nonconformal terms in the behavior of the deformations may be
affected by the $\alp$ and quantum effects, however.}
Generally speaking, because the KK 
wave functions are concentrated near the tip of the throat,
the deformations that interact most with KK modes 
are those that decay least (or grow most) with increasing $z$.

Previous studies of throat KK modes have only considered
decaying deformations because deformations that grow with $z$ could eventually
become $\mathcal{O}(1)$ and spoil the KS geometry.  
However, \cite{bdkkm} has pointed out that 
growing deformations are allowed as long as their amplitude is sufficiently
suppressed since the throat extends only a finite distance. They also 
identified such perturbations that would naturally be generated
by nonperturbative effects and therefore have suppressed amplitudes (these
deformations also preseve the supersymmetry of the background). In this 
paper, we study the decays of angular KK modes induced by these growing
deformations; we find that they are the dominant decay modes despite 
the suppression of their amplitude.  Section \ref{s:compare}  compares
our results to the previous literature.

As it happens, the growing deformations discussed in \cite{bdkkm} are
precisely the breathing mode of the $T^{1,1}$,
which is accompanied by a fluctuation of the warp factor (which we have
checked does not change the order-of-magnitude estimates we present later).  
There is also 
a similar deformation of the 4-form potential, but we focus on the
metric breathing mode. This deformation is in fact forbidden in the 
tree level supergravity \cite{GKP}, so it must be sourced by nonperturbative
or string physics, which suppresses its amplitude.
For a small amplitude 
deformation $\Delta \Gamma$ of the breathing mode $\Gamma$, 
the deformation takes the form
\eq{def1}{\Delta\Gamma(z,\theta) \approx w^4 e^{2kz} Y_{(j,l,r)}(\theta)}
for $(j,l,r)=(1,0,0)$ or $(0,1,0)$,
where $Y$ is the scalar angular harmonic function on $T^{1,1}$ \cite{bdkkm}.
Henceforth, we  use a capital $\Delta$ to represent background
deformations (as opposed to lower case $\delta$ for dynamical fluctuations).

In addition, ref.\ \cite{bdkkm} discusses deformations of the form
(\ref{gtdef})  with charge $(1/2,1/2,\pm 1)$, which have radial form
$\Delta\t\Gamma\sim w^4 e^{5kz/2}$. 
While we focus on the deformation (\ref{def1}), it
is possible that the lightest charged state decays by neutralizing its
charge against this
second deformation.  In  that case, as we discuss in more depth in
section \ref{s:decaymisc}, the decay rates may be enhanced by a factor
of $w^{-1}$.  However, this deformation can be forbidden by a
discrete symmetry of the CY.   In that case, if the lightest charged
KK mode has charge $(1/2,1/2,\pm 1)$, its decay rate could be
suppressed compared to our estimates due to the need for insertions
of multiple background deformations, for absorbing the angular
charges.  Since there are often multiple decay channels for the KK
modes, this situation may or may not be important in practice.

\subsubsection{Quadratic mixings}

The background deformation $\Delta\Gamma$, since it carries angular charge,
induces quadratic mixings between the charged and uncharged KK modes. 
While a careful accounting would include mixing in kinetic terms, we
restrict our study to off-diagonal mass terms.  From the 4D point of view,
these couplings control the oscillation of $\dgs$ into $\delta\Gamma$.  

A brief note on philosophy: we are treating the background deformation 
$\Delta\Gamma$ as inducing a number of interactions, including quadratic 
terms, and we work in the interaction picture.  This means we use the 
eigenmodes of the unperturbed background and introduce mixing diagrams when
calculating decay amplitudes.  An alternate approach, espoused by \cite{dkp},
is to diagonalize the quadratic action, including the new mixings, and
then work out the interactions of the new eigenmodes.  These approaches are
equivalent, as long as the mixing is a small perturbation (otherwise
one must resum the series of propagator insertions).  

The appropriate potential is then \cite{gm}
\eq{potential1}{ \frac{\L}{M_p^2} = -\frac{1}{2V_w} \int d^6y \sqrt{\t g}\t R
+\frac{g_s}{24V_w}\int d^6y\sqrt{\t g}\,e^{4A}G_{mnp}
(\b G-i\t\star_6 \b G)^{\widetilde{mnp}} \ ,}
where $\t g$ and $\t R$ refer to the metric (\ref{gtdef}).\footnote{As 
mentioned above, we are ignoring terms that depend on the spacetime 
derivatives of the fluctuations since this does not affect the
order of magnitude estimates.  There are also other $\alp$ or nonperturbative
corrections, but these also generically do not affect our estimates.} 
Both pieces of
the potential contribute terms of the form $\Delta\Gamma\,\delta\Gamma\,\dgs$; 
the first because the deformed fluctuating metric is no longer Ricci flat,
and the second because the deformed fluctuating Hodge star
$\t\star_6$ scales differently
for different numbers of angular legs on $\bar G_3$.  Specifically, the 6D 
Ricci scalar of (\ref{gtdef}) is 
\eq{riccitilde}{\t R = e^{2kz}\left[20k^2\left(e^{-2\Gamma}-1\right)+50k
\del_z\Gamma-5\del_z^2\Gamma -30(\del_z\Gamma)^2-8k^2e^{-2\Gamma}\hat\Del^2
\Gamma+3k^2e^{-2\Gamma}\left(\hat\Del\Gamma\right)^{\hat 2}\right]\ ,}
from which we can take the appropriate terms by defining $\Gamma=\Delta\Gamma
+\delta\Gamma+\dgs$.  To find the contribution from the flux, we note that
$\b G=i\t\star_6\b G$ on the conifold, so 
\eq{starG}{(\b G-i\t\star_6 \b G)^{\widetilde{z\theta\phi}}=
k^4 e^{6kz} e^{-4\Gamma}\left(1-e^{-\Gamma}\right)\b G^{z\hat\theta\hat\phi}
\ , \ \ (\b G-i\t\star_6 \b G)^{\widetilde{\theta\phi\psi}}=
k^6 e^{6kz} e^{-6\Gamma}\left(1-e^{\Gamma}\right)
\b G^{\hat\theta\hat\phi\hat\psi}\ .}  
Following, for example, \cite{fm}, we also note that the flux in the KS throat
is approximately constant:
\eq{fluxKS}{G_{z\theta\phi}\approx \frac{G}{\sqrt{6}k}\ ,\ \ 
G_{\theta\phi\psi}\approx \frac{G}{\sqrt{2}k^2}\ ,}
where $G$ is a complex constant related to the number of RR and NSNS 
flux quanta.

Putting everything together, including the normalized wave functions
(\ref{gammaflucts}), we find a mixing term between the charged and uncharged
scalars.  Since the wave functions are localized in the throat, the 
integration over $y$ only runs over the throat and is dominated by the 
tip end.  Furthermore, we find that the parameter $\nu$ cancels out.
The mixing term is
\eq{mixing1}{{\L}_{\gamma\gamma^\star} \approx N k^2 w^4\gamma(x)
\gamma^\star(x)\ ,}
where $N$ is a constant ranging from order unity to order 100, depending on
the flux.  Since the masses of $\gamma$ and $\gamma^\star$ are of order $wk$,
this induces a mixing angle of order $w^2$ between the Lagrangian fields
and the mass eigenstates.

It is similarly possible to calculate mixing of $\dgs$ with the universal 
volume modulus $c$
of the compactification (which is described in \cite{gm,ftud}).
While there are derivative couplings, we focus on the coupling
induced by the flux term in (\ref{potential1}).  Specifically, to first order
in the modulus, the warp factor becomes $e^{4A}+e^{8A}c/M_p$; $\dgs$ and 
$\Delta\Gamma$ each appear linearly through (\ref{starG}).
For $\nu^\star<4$, which 
we assume, the integral is dominated by the UV, so we find
\eq{mixing2}{{\L}_{\gamma^\star c} \approx \frac{M_s^4}{M_p k} 
\frac{w^{5+\nu^\star}}{4-\nu^\star} \gamma^\star c}
for a canonically normalized modulus.  We should note that mixing between
KK modes and the modulus is not possible before the modulus is stabilized 
(due to orthogonality of the metric excitations); therefore, any mixing
between the uncharged KK mode $\delta\Gamma$ and K\"ahler moduli must occur
at least at quadratic order in the deformation $\Delta\Gamma$ due to the
isometry.  The mixing of (\ref{mixing2}) is linear in $\Delta\Gamma$, as 
we have seen; this is of course necessary to keep the angular integrals 
from vanishing.

\subsubsection{Trilinear couplings}\label{s:axionints}

One type of decay we  want to consider is the decay of our DM candidate
to moduli of the compactification, since the moduli should be lighter than
KK modes.  Some of these decays exist in all the models we consider, so
they give an upper bound on the lifetime of the KK dark matter.  Here
we collect the cubic interactions that we study.

We first consider cubic couplings with moduli, taking the universal volume
modulus as a proxy for all the moduli.  To start with, we note that any
coupling linear in the KK modes and quadratic in moduli must involve at least
one power of the deformation $\Delta\Gamma$.  Any dimension-3 term of order 
$(\Delta\Gamma)^0$ would exist even in the absence of moduli stabilization, 
and, since the moduli can by definition take any expectation values in that
case, such terms would generate tadpoles for the KK modes.  Hence, 
$\dgs$ has no dimension-3 couplings at lowest order to one uncharged
and one charged modulus.  Similarly, $\delta\Gamma$ has no zeroth order 
dimension-3 couplings to two uncharged or two charged moduli.  This same 
argument applies, of course, to axions and to zero-derivative terms in the 
potential.

On the other hand, the deformation $\Delta\Gamma$ is sourced by the 10D
effect that stabilizes the moduli.  Therefore, there are dimension-3 
interactions between, for example, $\dgs$ and two universal volume moduli,
which is suppressed by one power of $\Delta\Gamma$.  This interaction 
also follows from the flux term of (\ref{potential1}) and gives
\eq{gcc}{{\L}_{\gamma^\star c^2}\approx \frac{M_s^4}{M_p^2k}
\frac{w^{5+\nu^\star}}{8-\nu^\star}\gamma^\star c^2}
for canonically normalized $c$. 

We also consider dimension-5 cubic couplings, specifically those with three
scalars and two derivatives.  The simplest of these is actually a 
self-coupling of the volume modulus, which arises due to its nontrivial
kinetic term (see \cite{ftud}).  In position and momentum space, it is
\eq{volself}{{\L}_{c^3}\approx \frac{1}{M_p} c(\del c)^2\approx -
\frac{p_1\cdot p_2}{M_p} c^3\ .}

There are also direct dimension-5 couplings between the KK modes and the
moduli sector, which are easier to estimate for the axions.
As an example, consider the coupling of 
$\delta\Gamma$ to two universal axions; the $T^{1,1}$ breathing mode appears
in both the metric determinant and in the contraction of 4-form indices.
If we focus on the $z\theta\phi\psi$ component of the universal axion, we
can estimate this interaction as
\bea{\L}_{\textnormal{\scriptsize axion}}&\approx& M_s^8\int d^6y \sqrt{\t g}\, 
e^{4A}\delta\Gamma(x,y)\del_\mu C_{z\theta\phi\psi}\del^\mu 
C^{\widetilde{z\theta\phi\psi}}\nonumber\\
&\approx& \int_0^{z_0} dz\int d^5\theta\sqrt{\hat g}\, k^6\delta\Gamma(x,y)
e^{-2kz}\del_\mu C_{z\theta\phi\psi}\del^\mu C_z{}^{\widehat{\theta\phi\psi}}
\ .\label{gammaaxax}
\eea
In the first line of (\ref{gammaaxax}), the warp factors enter in the 
external and internal metric determinants, the external metric contraction,
and the internal metric contractions; in the second line, additional 
exponentials of $kz$ come from $\tilde g_{mn}$ in the determinant and
contractions.  Other $z$ dependence is still implicit in the wavefunctions 
of the fluctuations.
There are, of course, other components, but those give order one corrections
to the interaction.
Due to the localization of the KK mode near $z_0$, we can do the integral just
over the throat; note that this leads to a suppression in the interaction
because the axion wave function decays in the throat.
In terms of canonically normalized $\gamma(x)$ and axions $a(x)$, the
interaction term in momentum space is
\eq{gammaaxax2}{{\L}_{\gamma a^2}\approx \frac{M_s^4}{M_p^2}
\frac{p_1\cdot p_2}{(8-\nu)k^3}
w^{1+\nu}\gamma a^2\ .}
Here $p_{1,2}$ are the momenta of the two axions.  We have assumed that
$\nu<8$, which means the integral is dominated by the UV end of the throat;
in the opposite case, take $w^{1+\nu}\to w^9$.  Here, and in the following,
we have replaced the warped volume with the string scale using 
(\ref{vwarped}).

There is also a direct cubic coupling between $\dgs$ and two universal 
axions induced by $\Delta\Gamma$, which comes from the same kinetic term.  
We have the coupling
\eq{gsaxax}{{\L}_{\gamma^\star a^2}\approx \frac{M_s^4}{M_p^2}
\frac{p_1\cdot p_2}{(6-\nu^\star)k^3}
w^{5+\nu^\star}\gamma^\star a^2
\ ,} assuming $\nu^\star <6$ (which seems likely, as mentioned in section
\ref{s:kkmodes}).  If $\nu^\star>6$, larger than our usual assumption, 
we take $w^{5+\nu^\star}\to w^{11}$.

In addition, we discussed in section \ref{s:moduli} the possibility of 
moduli with angular quantum numbers in the throat.  If these exist, our KK 
DM candidate can decay directly into one of these axions (plus a universal 
axion).  We also mentioned two possibilities for the wave function of these
charged axions, which we denote $a^\star$. 
The first is that the wave function
decays as $e^{-4kz}$ in the throat, like the universal axion's wave function.
In that case, the interaction term takes the form of (\ref{gammaaxax2}),
with $\nu\to\nu^\star$, $\gamma\to\gamma^\star$, and $a^2\to aa^\star$.  The 
other extreme possibility, suggested by the $AdS_5$ dimensional
reduction, is that the wave function of $a^\star$ is constant
in the throat.  In that case, we find
\eq{gsaxaxt}{{\L}_{\gamma^\star a a^\star}\approx \frac{M_s^4}{M_p^2}
\frac{p_1\cdot p_2}{(4-\nu)k^3}
w^{1+\nu^\star}\gamma^\star aa^\star}
for $\nu^\star <4$; take $w^{1+\nu^\star}\to w^5$ for $\nu^\star>4$.
Note that both of these possible wave functions give the same interaction for
$\nu^\star\leq 4$.

We may further consider decays to spacetime
gravitons $\dgs\to h_{\mu\nu}^2$. However, \cite{chentye,dkp} showed using
orthogonality of wave functions that this type of amplitude vanishes when 
$\dgs$ is replaced by a KK graviton.  We can generalize this result
to show that no field can decay to gravitons.
 Suppose
the 1PI effective action of our theory in Einstein frame takes the form
$S=S_{EH}+S_m(g,\Phi)$, where $\Phi$ are all matter fields, 
$S_{EH}$ is the usual Einstein-Hilbert 
action and that all backgrounds are covariantly constant.  Then all the 
terms in the action that are first order in any fluctuation $\delta\Phi$
take the form 
\eq{gravargument}{-\int d^4x\,\sqrt{-g}\frac{\delta V}{\delta\Phi}\delta\Phi
=0\ ,}
since we evaluate $\delta V/\delta\Phi$ on the background.  Therefore, there
are no couplings of the form $\Phi h^N$.  
Similar couplings are allowed, however,
when we include higher derivative corrections to the gravitational action,
since we cannot rule out terms of the form $\Phi (R_{\mu\nu\lambda\rho})^N$
for $N\geq 2$ (we can rule out $N=1$
by converting to Einstein frame for $S_{EH}$).\footnote{Interestingly, 
time-varying backgrounds, such as quintessence, or other Lorentz breaking
backgrounds can induce interactions as well.  
However, these are unimportant for us 
because such interactions are suppressed by the Hubble scale and are further
not present for massive fields like KK modes.}
The simplest trilinear vertex is schematically 
\eq{gravvertex}{\L_{\Phi h^2}\approx \lambda\Phi\del^4 h^2\ , \ \ 
\lambda\lesssim \frac{1}{m_\Phi M_p^2}\ ,}
where the limit on the coupling $\lambda$ is saturated when the mass of
the decaying particle is the same as the cutoff of the effective field
theory.  It seems possible that this coupling is further suppressed by 
powers of the warp factor $w$, but we will take (\ref{gravvertex}) as an upper
bound.

\subsection{Brane interactions}\label{s:braneints}

Now we turn to trilinear couplings of KK modes with brane fields.

\subsubsection{D3-brane Standard Model}

We first examine models in which the SM lives on D3-branes at the tip of the 
throat.  We consider two types of couplings.
First, suppose that the position of the D3-brane on the angular dimensions 
at the tip of the throat completely breaks the isometries associated with the
charge of the lightest charged KK mode.  Of course, a single pointlike
brane cannot break the entire isometry of the throat, but the KK mode might
have motion outside the unbroken isometry group.  In practical terms, the
angular wave function of $\dgs$ does not vanish at the position of a D3-brane
that breaks the appropriate isometries.

There is first a dimension-5 coupling between $\dgs$ and the brane scalars,
which follows from the scaling of the angular components $\t g_{mn}$ under 
fluctuations of the $T^{1,1}$ breathing mode. This is
\eq{D3kinetic2}{\L_{\gamma^\star\phi^2}\ \approx\  
\frac{\mu_3}{g_s}\frac{w^2}{k^2}\dgs(x,y_0)\del_\mu\phi
\del^\mu\phi\ \approx\  k^3\frac{p_1\cdot p_2}{M_s^4 w}\gamma^\star\phi^2\ ,}
where we have converted to canonically normalized scalars 
using (\ref{D3kinetic3}) after the second
approximate equality and used the fact that $e^{(2+\nu^\star)k z_0} = 
w^{-(2+\nu_\star)}$ for a brane at the  tip of the throat.
We have assumed that the angular part of the $\dgs$ wave function is of 
order unity.  As in the axion interactions of section \ref{s:axionints},
$p_{1,2}$ are the momenta of the brane scalars. For $k\lesssim M_s$, this 
is suppressed by the warped string scale $w M_s$.

Similarly, there is a Yukawa coupling with the brane fermions.  
From \cite{hep-th/0202118} (see also
\cite{hep-th/0311241,hep-th/0312232}), the fermionic action on a D3-brane
includes the terms
\eq{D3yukawa}{\L_{\Theta_3}\approx g_s w \,\b\Theta_3\Gamma^{mnp}
\Theta_3\,\textnormal{Re} \left(iG-\t\star_6G\right)_{mnp}\ ,}
where $\Gamma^m$ is the Dirac matrix associated with the full compact metric
including warp factors, and $\Gamma^{mnp}$ is the antisymmetrized
product of $\Gamma^m\Gamma^n\Gamma^p$.  We have normalized the 
fermions canonically.  The flux component in (\ref{D3yukawa}) is 
\eq{fluxaisd}{\textnormal{Re} \left(iG-\t\star_6G\right)_{z\theta\phi}=
g_{\mathrm{s}}^{-1}\dgs H_{z\theta\phi}\ , \ \ \textnormal{Re} 
\left(iG-\t\star_6G\right)_{\theta\phi\psi}=-g_{\mathrm{s}}^{-1}\dgs 
H_{\theta\phi\psi}\ .}
In the usual KS solution, the NSNS flux $H$ has a leg on $z$, but the 
parametrics of the interaction are the same for either component, and some
4D fermions  couple to $\dgs$ in either case.  Then  
\eq{D3yukawa2}{\L_{\gamma^\star\bar\Theta_3\Theta_3}\approx
wk\textnormal{Re}(iG) \dgs(x,y_0)\b\Theta_3\Theta_3\approx
\left(\frac{k}{M_s}\right)^4 \gamma^\star\b\Theta_3\Theta_3 }
with canonically normalized $\gamma^\star(x)$.

The second class of couplings is the one obtained by substituting 
$\delta\Gamma$ for $\dgs\!$.  In fact, the scaling and dependence on all 
variables is the same as in (\ref{D3kinetic2}) and (\ref{D3yukawa2}).
These couplings are important in the case that $\dgs$ vanishes at the 
position of the D3-brane, either because the brane does not break the 
appropriate isometry (including the case that the $\dgs$ KK mode has a 
vanishing wave function
at $z_0$ due to a centrifugal barrier). It should be noted that
the volume modulus has a suppressed coupling to D3-branes due to the small
warp factor at the tip, so we do not consider this coupling.

\subsubsection{D7-brane Standard Model}\label{s:d7coupling}

We now turn to models with the Standard Model living on D7-branes in the KS
throat.  Similarly to the D3-brane Standard Model, 
we are interested in two sets
of interactions, those directly coupling $\dgs$ to the D7-brane fields 
and those coupling $\delta\Gamma$ to the brane fields.  A point of interest,
of course, is whether the D7-brane preserves enough isometries to forbid 
some of the direct couplings.  While a particular D7-brane
embedding realizes only one of these possibilities, 
we consider both cases since more general embeddings probe them both.

In either case, the interaction arises once again because the DBI action
of the brane is given by the induced metric on the brane.  Including 
the $T^{1,1}$ breathing mode in the metric (\ref{d7pullback}), 
the breathing mode should appear in $\t g_{\chi\b\chi}$ as well as in the 
longitudinal angular components (some of the $y^i$ directions) within
the KS throat.  This is simply because $\chi$ is approximately an angular
coordinate.

Suppose first that the D7-brane breaks enough isometries that the 
world-volume angular integral of $\dgs$ does not vanish.  In this case, the
vertex is given by the integral (\ref{d7kinetic}) with $\dgs$ inserted
and canonical normalization taken for all fields.  This is approximately
\eq{D7direct}{\L_{\gamma^\star\chi^2}\approx \left(\frac{M_s}{M_p}\right)^{4/3}
\frac{w}{k}
\left(\frac{w}{w_1}\right)^{\nu^\star}
p_1\cdot p_2\, \gamma^\star |\chi|^2 \ .}

Even in the case that the D7-brane maintains the isometries, there can 
be a direct coupling of $\dgs$ to the brane along with the indirect coupling
through a $\delta\Gamma$ intermediate state.  Both of these are summarized
in the term
\eq{d7couplings}{\L_\chi\approx \int_0^{z_1} dz\int 
d^3\theta\sqrt{\hat g}e^{-2kz}\left(\delta\Gamma+\Delta\Gamma\dgs\right)
\del_\mu\chi\del^\mu\b\chi\ , }
where we have again used canonically normalized $\chi$.  These couplings
integrate to
\eq{d7couplings2}{\L_{\gamma\chi^2}\approx\left(\frac{M_s}{M_p}\right)^{4/3}
\frac{w}{k}
\left(\frac{w}{w_1}\right)^{\nu} 
p_1\cdot p_2 \,\gamma |\chi|^2
\ , \ \ \L_{\gamma^\star\chi^2}\approx \left(\frac{M_s}{M_p}\right)^{4/3}
\frac{w^3}{k}
\left(\frac{w}{w_1}\right)^{2+\nu^\star} 
p_1\cdot p_2 \,\gamma^\star |\chi|^2}
with canonical normalization. Moreover, the universal volume modulus 
couples similarly to the brane scalars; in this case, we can just read the
coupling from (\ref{d7kinetic}) through the shift of the warp factor 
(\ref{univvol}).  If we approximate $V_{4w}$ as below eq.\ (\ref{d7kinetic})
 which is
reasonable, since we work in coordinates where the warp factor is nearly
unity in the bulk CY, we immediately find
\eq{d7couplings3}{\L_{c\chi^2} \approx \frac{p_1\cdot p_2}{M_p} c|\chi|^2\ .}

We would also like to consider Yukawa interactions with the D7-brane fermions.
Unfortunately, the full couplings of D7-brane fermions to supergravity 
modes are not known in the presence of 3-form flux and general warp
factors.\footnote{In particular, note that \cite{hep-th/0311241} specifically
assumes that the warp factor is constant parallel to the brane; this choice
also obscures where the warp factor would appear implicitly.}  In 
principle, the Yukawa coupling can be determined by expanding the action
given in \cite{hep-th/0306066}, but that is beyond the scope of this paper.
However, 3-form flux can induce Yukawa interactions \cite{hep-th/0311241},
and we can estimate the Yukawa couplings simply, as follows.  To
order of magnitude, the D7-brane fermions and scalars should couple similarly
to the closed string degrees of freedom, but the dimension-5 operators
should be suppressed by a single power of the cutoff of the D7-brane
effective field theory.  This scale is approximately $w_1k$, so we can 
estimate the corresponding Yukawa couplings by multiplying, for 
example, (\ref{d7couplings2})
 by $w_1k$ (and removing the $p_1\cdot p_2$ factor):
\eq{d7couplings4}{\L_{\gamma\b\Theta_7\Theta_7}
\approx\left(\frac{M_s}{M_p}\right)^{4/3}w^2
\left(\frac{w}{w_1}\right)^{-1+\nu} \,\gamma \b\Theta_7\Theta_7
\ , \ \
\L_{\gamma^\star\bar\Theta_7\Theta_7}\ \approx \
\left(\frac{M_s}{M_p}\right)^{4/3}w^4
\left(\frac{w}{w_1}\right)^{1+\nu^\star}
\, \gamma^\star \bar\Theta_7\Theta_7 \ .}
  Note that the D3-brane couplings satisfy this 
relation as well, with cutoff $wk$.

\section{Decay rates}\label{s:decay}

In this section, we assemble the interactions into decay rates for
the DM candidates.  There are several different possible rates, depending on
specifics of the Standard Model construction, the CY manifold, and the 
specific DM candidate.

When making numerical estimates for the decay rates, we use the following
parameter values appropriate for $\tev$ scale dark matter, as indicated
in the decay rates below: $M_s\approx M_{GUT}\approx 10^{16}\ \gev$, 
$k\approx M_s$, $w\approx 10^{-13}$, and $w_1\approx 10^{-4}$.  The
value of $w_1$ assumes an intermediate scale for the D7-brane SM with the 
SM hierarchy generated by supersymmetry breaking (this is the same
scale as discussed in \cite{d7susy}, for example).

\subsection{Decays to supergravity modes}\label{s:sugradecay}
We first consider decays to light supergravity modes, specifically moduli
and their axion partners.  After moduli stabilization, we expect that these
are actually heavy compared to the lightest Standard Model particles; 
assuming reasonable couplings, these supergravity modes can decay quickly
to SM states.  On the other hand, if the light supergravity states couple
only weakly to the SM, they may constitute a different candidate DM sector,
along the lines suggested in ref.\ \cite{arXiv:0906.3346}.  We leave this 
possibility for other work, as we are interested in the lifetime of the 
charged KK mode sector as DM candidates.

It turns out that both dimension-3 and dimension-5 interactions are important
for decays to light supergravity modes, so we consider both here.
We begin with decays to the universal volume modulus, which should be 
representative of decays to other uncharged moduli.  This decay can proceed
through two channels: first, our DM candidate $\dgs$ has a direct dimension-3
coupling (\ref{gcc}) with two moduli; second, $\dgs$ can oscillate into a 
volume modulus through (\ref{mixing2}), which then decays via the 
dimension-5 self-coupling (\ref{volself}).
The amplitudes for both of these processes are parametrically similar,
and they yield a decay rate of order (using 
$\Gamma\sim m_\star^{-1} |\mathcal{M}|^2$ for a scalar of mass $m_\star$ 
decaying into much lighter particles)
\bea
\Gamma(\gamma^\star\to cc) &\approx& \frac{M_s^8}{M_p^4 k^3} w^{9+2\nu^\star}
\nonumber\\ &\approx& 10^{-113-26\nu^\star} \gev  \times f(-3,5,9+2\nu^\star)
\nonumber\\
&\approx& 10^{-89-26\nu^\star} \textnormal{s}^{-1} 
\times f(-3,5,9+2\nu^\star)\ .\label{voldecay}\eea
where
\bea
	f(i,j,k) = 
\left(\frac{k}{M_s}\right)^i 
\left(\frac{M_s}{10^{16}\ \gev}\right)^{j} (10^{13} w)^{k}\ .
\eea

Although the two diagrams do not cancel to the approximations we have used,
we note that they could possibly cancel in a more complete 
calculation, even though there is no apparent symmetry-based reason 
that they should.
In that case, the decay proceeds through a dimension-5 coupling, which,
as we now show, may be competetive with (\ref{voldecay}) in some 
circumstances.

We estimate these decay rates by considering the dimension-5 couplings 
to the universal axion given in (\ref{gammaaxax},\ref{gammaaxax2}).  The
amplitudes involving these vertices (and quadratic mixing when appropriate)
have different powers in the warp factor, so one or the other process 
dominates in different models.  For $\nu<\nu^\star+2$, the process with 
$\gamma$-$\gamma^\star$ mixing (\ref{mixing1}), 
intermediate $\gamma$ state and vertex (\ref{gammaaxax2}) dominates and
yields a decay rate which is equal to that which we would get from 
$\gamma\to 2a$,
but suppressed by the mixing angle squared $\theta^2\sim w^4$.  The result is
\bea 
\Gamma(\gamma^\star\to aa) \ = \ 
\Gamma(\gamma^\star\to cc)\ \textnormal{with}\ \nu^\star\to\nu
\label{axiondecay1}\eea
which indeed is larger than (\ref{voldecay}) when $\nu<\nu^\star$.
(In the intermediate state propagator, we have assumed that the mass squared
of both $\dgs$ and $\delta\Gamma$ are of order $wk$, as is the difference,
and we continue to make this assumption throughout; furthermore, 
although we should sum over the entire radial KK tower, the sum is convergent
and should not affect order of magnitude estimates.)
For $\nu>\nu^\star+2$, the direct $\gamma^\star\to 2a$
 decay amplitude (\ref{gsaxax}) dominates, and we find
\bea
\Gamma(\gamma^\star\to aa) &\approx& \frac{M_s^8}{M_p^4 k^3} w^{13+2\nu^\star}
\nonumber\\ &\approx& 10^{-165-26\nu^\star} \gev \times f({-3,5,13+2\nu^\star})
\nonumber\\
&\approx& 10^{-141-26\nu^\star} \textnormal{s}^{-1}
\times f({-3,5,13+2\nu^\star})\ .\label{axiondecay2}\eea

Recall that the charged KK state can decay 
to an axion with angular charge in the throat and an uncharged axion 
(such as the universal axion) through the direct dimension-5 coupling
(\ref{gsaxaxt}), assuming such a charged modulus and axion exist.  In this
case, we find a much faster decay channel,
\bea
\Gamma(\gamma^\star\to a a^\star) &\approx& 
\frac{M_s^8}{M_p^4 k^3} w^{5+2\nu^\star}\nonumber\\
&\approx& 10^{-61-26\nu^\star} \gev \times f({-3,5,5+2\nu^\star})\nonumber\\
&\approx& 10^{-37-26\nu^\star} \textnormal{s}^{-1}\times f({-3,5,5+2\nu^\star})
\ .\label{axiondecay3}\eea
Despite the fact that this is by far the fastest of the decay rates we have 
found to the modulus/axion sector, the lifetime 
is still many orders of magnitude longer than the age of the 
universe.

Finally, we can give an upper bound for decay to 4D gravitons using the 
inequality in eq.\  (\ref{gravvertex}) for the dimension 7 coupling
$\lambda$.  Taking 
$m_\Phi\sim wk$ and also replacing the graviton derivatives by $\del\to wk$,
we find a decay rate
\bea\Gamma(\gamma^\star\to hh) &\lesssim &\frac{w^5k^5}{M_p^4} \approx
10^{-61}\ \gev\ \times\ f(5,5,5)\nonumber\\
&\approx& 10^{-37}\ \textnormal{s}^{-1}\ \times f(5,5,5)\ .
\label{gravdecay}\eea
This is much faster than the other decays to bulk supergravity modes
(except for (\ref{axiondecay3}) when $\nu^\star\sim 0$), but it still
results in a lifetime which is   much much longer than the age of the
universe. Notice that this is an upper bound on the rate of
decay into gravitons.

\subsection{Decays to D3-brane Standard Model}

Now we turn to decays to SM particles with the SM living on D3-branes at
$z_0$.  There are two possibilities for these decays: either $\dgs$ has 
a vanishing wave function at the position of the brane (for symmetry reasons
such as centrifugal barrier) or not.  We begin with the case in which
there is a nonvanishing direct coupling of $\dgs$ to the brane.

\subsubsection{Direct decay through symmetry breaking}

In this case, the direct interactions of $\dgs$ with the brane fields are 
given by (\ref{D3kinetic2}) and (\ref{D3yukawa2}) for couplings to the 
brane scalars and fermions respectively.  
For scalars, the rate simply follows from the usual formula 
$\Gamma\sim m_\star^{-1} |\mathcal{M}|^2$, where the Feynman diagram for the
decay is just the vertex.  Recall that the 
inner product of momenta is $p_1\cdot p_2\sim wk$ for decay products much
lighter than the DM candidates.  Then the decay rate is
\eq{d3rate1}{\Gamma(\gamma^\star\to \phi\phi)\ 
\approx\ \frac{wk^9}{M_s^8}\ \approx\ 1\ \tev \times
f(9,1,1)
\ \approx\  10^{27}\ \textnormal{s}^{-1} \times
f(9,1,1)\ . }
This is extremely rapid ($\tev$ scale) and would obviously
rule out these angular KK modes as DM candidates if it is not forbidden.

The decay rate to fermions also follows from a single vertex; the only 
additional element is the trace over external spinors, which simply gives
$\textnormal{tr}[\slashed{p}_1\slashed{p}_2]\sim p_1\cdot p_2$.  Working
through the algebra, we find the same decay rate (\ref{d3rate1}) to order
of magnitude even though the scalar decay proceeded via a dimension-5 
operator; that is, $\Gamma(\gamma^\star\to \b\Theta_3\Theta_3)
\approx\Gamma(\gamma^\star\to \phi\phi)$.  In retrospect this is
understandable because the DM mass is similar to 
the natural cutoff of the D3-brane theory.

It is noteworthy that there exists a simple dynamical mechanism 
whereby these fast
decay channels can be forbidden.  Namely, the breaking of angular 
isometries implies
that the brane modulus is stabilized on some particular 
subspace of the $T^{1,1}$
manifold.  It is plausible that the potential is minimized for 
angles where wave function 
$Y(\theta_i)$ vanishes.  As a specific example, \cite{firtye} showed
that states charged under one of the $SU(2)$ factors face a centrifugal
barrier due to the shrinking of the $S^2$ at the tip of the throat, so the
wave function vanishes at the tip.  This is related to the fact that this 
$SU(2)$ factor rotates the $\mathbb{R}^3$ formed by the shrinking $S^2$ and
the radial direction, a symmetry unbroken by the D3-brane at the tip, the
origin of that $\mathbb{R}^3$.  Then the brane position does not 
break the symmetry generator under which  the KK mode has charge.

\subsubsection{Indirect decay only}

If there is a dynamical mechanism such as the one mentioned above, 
then the lightest
charged state has a vanishing wave function on the D3-brane.  
Alternatively, it could
happen that  the DM candidate has angular momentum  on the 
2-cycle that shrinks at the
throat tip; this is in fact a variation of the previous option, as the D3-brane
position on the $S^3$ at the tip does not break angular symmetries 
of the (shrinking) $S^2$.

In such cases the decay amplitude includes mixing of $\gamma^\star$ 
with $\gamma$, a $\gamma$ propagator, and the vertex of $\gamma$ on the brane.
As in the case of the direct decay, the decay rates into scalars and
fermions are parametrically similar, and the common rate is
\bea
\label{d3rate2}\Gamma(\gamma^\star\to\phi\phi)&\approx& \Gamma(\gamma^\star\to
\b\Theta_3\Theta_3)\ \approx\  \frac{w^5k^9}{M_s^8}\nonumber\\ 
&\approx& \ 
10^{-49}\ \gev \times f(9,1,5)\ \approx 
\ 10^{-25}\ \textnormal{s}^{-1}\times f(9,1,5) \ .
\eea

This rate is particularly interesting; for the fiducial values we have 
chosen, it is close to that needed to explain the high-energy 
positron excess seen by PAMELA and electron-positron flux excess seen
by Fermi, ATIC, HESS, and other experiments \cite{decays1,decays2}.
Recall that in a realistic embedding of the SM on the brane,
$\Theta_3$ would represent SM fermions, gaugini, or Higgsino,
and $\phi$ would be the Higgs boson or sfermions of the MSSM.  
We have dropped some model-dependent factors such as flux quantum numbers 
which can 
shift this prediction by a few orders of magnitude, 
but one immediately sees from
(\ref{d3rate2}) that slight changes in the $\ads$ radius, warp
factor, or string scale can compensate for that.  One interesting
direction for future research would be to determine if 
this class of models can be arranged to provide leptophilic DM decays, 
which would be consistent with observation.  However, given the difficulty
of that task, it is perhaps more interesting and important to use this
rate to constrain D3-brane Standard Models.  The point is that any DM decay
to SM particles of this rate or above would be detected; therefore, if
a detailed thermal history yields an appropriate density of charged KK states,
certain throat geometry and flux configurations can be ruled out.  
Specifically, it may be possible to rule out some region of $w$, $k$, $M_s$
parameter space.

\subsection{Decays to D7-brane Standard Model}\label{s:d7decay}

The last category of decays we consider is to SM particles with the
SM on a D7-brane which extends some distance into the KS throat. 
Again, there  are the same two cases as for the D3-brane Standard
Model.

\subsubsection{Direct decay through symmetry breaking}

In this case, as before, the position of the D7-brane breaks the isometry
associated with the angular motion of the KK DM candidate.  Then
the world-volume integral of $\dgs$ does not vanish, so $\gamma^\star$ 
can decay by the interaction (\ref{D7direct}) to D7-brane scalars.  
Using that interaction, the rate comes to
\bea
\Gamma(\gamma^\star\to\b\chi\chi)
&\ \approx\ & w^5k\left(\frac{M_s}{M_p}\right)^{8/3} 
\left(\frac{w}{w_1}\right)^{2\nu^\star}\nonumber\\ &\approx& 
10^{-57-18\nu^\star}\ \gev \times f(1,11/3,5)\, g(2\nu^\star)
\nonumber\\
&\ \approx\ & 10^{-33-18\nu^\star}\ \textnormal{s}^{-1} 
\times f(1,11/3,5)\, g(2\nu^\star)
\ .\label{d7decay1}
\eea
where
$
	g(n) = \left(10^9 w/w_1\right)^n
$.
For small values $\nu^\star\gtrsim 0$ and large $w$ ($w\gtrsim 10^{-12}$), 
this decay mode comes close to being observable.  With care, some corner 
of parameter space could be ruled out for these models; this statement is
stronger for SM fermions, as we now see.

We can also estimate the decay to SM fermions in this case.  Using
the interaction (\ref{d7couplings4}) leads to the same rate as
 (\ref{d7decay1}) except rescaled by $(w_1/w)^2$.  Thus, we estimate the
decay to D7-brane fermions as
\bea
\Gamma(\gamma^\star\to \b\Theta_7\Theta_7)
&\ \approx\ & w^5 k \left(\frac{M_s}{M_p}\right)^{8/3} 
\left(\frac{w}{w_1}\right)^{-2+2\nu^\star}\nonumber\\ &\approx& 
10^{-39-18\nu^\star}\ \gev \times f(1,11/3,5)\, g(-2+2\nu^\star)
\nonumber\\
&\ \approx\ & 10^{-15-18\nu^\star}\ \textnormal{s}^{-1} 
\times f(1,11/3,5)\, g(-2+2\nu^\star)
\ .\label{d7decay2}
\eea
Again, this rate can become observable in the large $w$ region of parameter
space for $\nu^\star\lesssim 1$.  This model is a good candidate for 
future detailed study, including a careful calculation of the Yukawa coupling.

Notice that these decay rates are slower than for the corresponding case
of decay to a D3-brane SM.  The reason is that the wave function of the KK 
mode DM is localized near the tip of the throat and has maximal overlap with 
the D3-brane, while the D7-brane extends only a relatively short distance
into the throat where the KK mode wave function is small.

\subsubsection{Indirect decay only}

If the D7-brane does not break the appropriate symmetries of the lightest 
charged KK mode, then the decay amplitude involves the throat deformation
$\Delta\Gamma$ and, in some terms, intermediate states.  There are three
relevant diagrams: one with $\dgs$ scattering off $\Delta\Gamma$ on the brane,
one with a $\delta\Gamma$ intermediate state, and one with the universal
volume modulus as the intermediate state.  These three diagrams give 
parametrically different contributions to the decay amplitude, so they 
 dominate in different models.  

The diagram with the universal volume modulus (or other uncharged modulus)
as an intermediate state dominates the decay amplitude when $\nu^\star$ 
is particularly small or when the D7-brane does not stretch very far into
the throat ($w_1$ is large).  In fact, this process can occur even if the
D7-brane is localized entirely in the bulk CY, due to the fact that the
modulus wave function spreads throughout the entire compactification.
Putting together the quadratic mixing (\ref{mixing2}), the propagator 
discussed in section \ref{s:sugradecay} above, and the interaction 
(\ref{d7couplings3}), we find approximately
\bea
\Gamma(\gamma^\star\to\b\chi\chi)
&\approx& w^{9+2\nu^\star} \frac{M_s^8}{M_p^4 k^3}\nonumber\\
 &\approx& 
10^{-113-26\nu^\star} \ \gev \times f(-3,5,9+2\nu^\star) 
\nonumber\\
&\ \approx\ & 10^{-89-26\nu^\star}\ \textnormal{s}^{-1}
\times f(-3,5,9+2\nu^\star) \ .\label{d7decay3}
\eea

For longer throats with $\nu^\star<\nu+2$, the dimension-5 coupling of 
$\dgs$ to $\chi$ dominates.  In that case, the decay rate is
\bea\Gamma(\gamma^\star\to\b\chi\chi) 
&\ \approx\  & kw^9\left(\frac{w}{w_1}\right)^{4+2\nu^\star}
\left(\frac{M_s}{M_p}\right)^{8/3}\nonumber\\ &\approx& 
10^{-145-18\nu^\star}\ \gev \times f(1,11/3,9)\, g(4+2\nu^\star) 
\nonumber\\
&\ \approx\ & 10^{-121-18\nu^\star}\  \textnormal{s}^{-1} 
\times f(1,11/3,9)\, g(4+2\nu^\star) 
\ .\label{d7decay4}\eea

In other cases, the diagram with an intermediate uncharged KK mode
dominates the decay amplitude.  Then the decay rate is
\bea\Gamma(\gamma^\star\to\b\chi\chi) 
&\ \approx\  & kw^9\left(\frac{w}{w_1}\right)^{2\nu}
\left(\frac{M_s}{M_p}\right)^{8/3}\nonumber\\ &\approx&
10^{-109-18\nu}\ \gev\ \times f(1,11/3,9)\, g(2\nu) 
\nonumber\\
&\ \approx\ & 10^{-85-18\nu}\ \textnormal{s}^{-1}
 \times f(1,11/3,9)\, g(2\nu)\ .\label{d7decay5}\eea
Notice that all the implied lifetimes are much longer than the age
of the universe.

To find decay rates to D7-brane fermions, our estimates tell us to 
multiply (\ref{d7decay3},\ref{d7decay4},\ref{d7decay5}) by $(w_1/w)^2$.
For the fiducial values we have chosen, this is a factor of $10^{18}$,
which still leaves all the indirect decay rates much 
slower than the current Hubble rate.
In this estimate, the decay channel with the intermediate modulus now
depends on $w_1$; when the D7-brane does not extend into the KS throat
containing the DM candidate, set $w_1\to 1$.  However, $\gamma^\star$ and
$\gamma$ cannot couple to D7-branes that do not extend into the throat.

\subsection{Tunneling to other throats}\label{s:tunnel}

In the case that the SM branes are located outside the throat, our DM
candidates may decay to supergravity modes as discussed in \ref{s:sugradecay}
above.  Alternatively, though, there are possible decay channels to the
SM sector, even in other throats.  One possibility, which we have already
mentioned in the caes of a D7-brane SM, is that the KK mode mixes with
a modulus (the universal volume modulus, for example), which couples to 
branes throughout the compactification, albeit weakly.  This process is 
related to the discussion of \cite{arXiv:0801.4015} for uncharged KK modes
coupling to gravitons.

Another possibility is that KK modes may
tunnel through the bulk CY to couple to the SM degrees of freedom. 
Similar
tunneling rates have been discussed in the context of  reheating in
\cite{hep-th/0412040,ky,fmm,hep-th/0508229,firtye,chentye} (and more
generally in \cite{hep-th/0104239,hep-th/0106128}); there is some
support for the idea that the longest throats reheat the most.  This gives
a reason to consider the SM to live in the throat with our
DM candidates, but we consider other models to be comprehensive. 
Tunneling is closely tied to the
cosmological history of dark matter, including reheating and
thermalization, but we give a brief analysis of its importance in DM decays
here.  

We will adopt the estimate of \cite{chentye} for the tunneling 
amplitude, which is given by a mixing angle $\sim w_A^4$ between uncharged 
KK modes localized near the tip of throat $A$ and modes localized in
a second throat $B$.\footnote{Note that this is more suppressed than earlier
estimates based on \cite{hep-th/0104239,hep-th/0106128}; we favor the 
results of \cite{chentye} because that work glues the throat to the bulk
CY in a smooth fashion, more representative of the full 
compactification.  However,
as we indicate, tunneling in warped compactifications is not completely
understood and is worth revisiting.}\ \   
Note that this amplitude is independent of the length of the 
throat $B$ containing the final state particles.  However, there is as yet
no analysis of tunneling rates for states charged under approximate isometries
of throat $A$; since the final state is in a different throat with different
(or no) approximate isometries, the tunneling rate may be affected.  Without
performing a careful calculation, we consider two extremes: first, that
the tunneling amplitude is unaffected due to the fact that the bulk, which 
the particle tunnels through, breaks all isometries strongly, and, second,
that the charged KK mode must oscillate into an uncharged KK mode before
tunneling.  The latter eventuality suppresses the tunneling amplitude by an 
additional factor of $w_A^2$ (see (\ref{mixing1}) and the subsequent
discussion).

There are two major cases for the final states after tunneling.  If the 
KK spectra of the two throats match closely, tunneling can occur nearly 
on-shell.  In that case, the KK mode of throat $A$ oscillates into a KK mode
of throat $B$, which later decays to the SM.  The two extreme
possibilities for this tunneling rate are
\bea \Gamma(\gamma^\star_A\to {\rm KK}_B) &\approx& w_A^9 k_A\approx
10^{-101}\ \gev\ \times f(1,1,9)\nonumber\\
&\approx& 10^{-77}\ \textnormal{s}^{-1} \ \times f(1,1,9)\ \ \ 
\textbf{or}\nonumber\\
\Gamma(\gamma^\star_A\to {\rm KK}_B)&\approx & w_A^{13} k_A\approx
10^{-153}\ \gev\ \times f(1,1,13)\nonumber\\
&\approx& 10^{-129}\ \textnormal{s}^{-1} \ \times f(1,1,13)\ .\label{tunnel1}
\eea
Note, though, that this does not give the direct decay rate into 
light particles; the KK$_B$ lifetime could still be significant 
compared to particle
physics scales or even the age of the universe, though we assume that it is
short compared to the tunneling rate.

If the KK spectra of the two throats are very different, then the 
$B$-throat KK mode must be
treated as an off-shell intermediate state.  In this case, the decay 
amplitude takes the form
\eq{tunnel2}{\textnormal{tunneling mixing angle} 
\times\ {\rm KK}_B\ \textnormal{decay amplitude}\ ,}
perhaps with oscillation into an uncharged KK mode of throat $A$ before 
tunneling. As an example, consider direct decay to a D3-brane at the tip of 
throat $B$.  In that case, we find 
\eq{tunnel3}{\Gamma(\gamma^\star_A\to \phi_B\phi_B)\approx w_A^8 
\frac{w_B k_B^9}{M_s^8}\ \ \textnormal{or}\ \  w_A^{12}
\frac{w_B k_B^9}{M_s^8}\ .}
These rates clearly depend on the parameters of the $B$ throat.

\subsection{Decay rate generalizations}\label{s:decaymisc}

We conclude our presentation of decay rates with a few general comments.

First, we note that we have already, in a subtle way, presented the
decay rates of uncharged KK modes.  Any decay of $\gamma^\star$ that does not
require an insertion of $\Delta\Gamma$ corresponds to a similar decay of
$\gamma$.  For example, (\ref{axiondecay3}) gives a decay rate for 
$\gamma^\star\to a a^\star$, but the decay $\gamma\to aa$ is parametrically
identical.  Similarly, the couplings of $\gamma^\star$ to isometry-breaking
branes are identical to the couplings of $\gamma$ to any brane (taking
$\nu^\star\to \nu$ when appropriate).  Therefore, we can immediately see
that an uncharged KK mode in a $\tev$ scale throat with a D3-brane at the tip
will decay with a $\tev$ scale rate.

We can also give some simple conversion rules for our decay rates in the
case that the lightest charged KK state carries a different angular charge.
As mentioned in section \ref{s:breaking}, the small perturbation 
$\Delta\t\Gamma$ with charge $(1/2,1/2,1)$ grows as $e^{5kz/2}$ in the throat.
If the lightest charged KK mode falls in this charge sector, then 
decay amplitudes involving an insertion of $\Delta\t\Gamma$ are boosted by a 
factor $w^{-1}$, assuming that the integrals done in the dimensional 
reduction are dominated by large $z$.  These include the 
$\gamma^\star$--$\gamma$ mixing (\ref{mixing1}), the $\gamma^\star$--$c$ 
mixing (\ref{mixing2}) for $\nu^\star>7/2$, the 
$\gamma^\star$--$a$--$a^\star$ 
coupling (\ref{gsaxaxt}) for $\nu^\star>7/2$ (or other $\gamma^\star$--axion
couplings at large $\nu^\star$), and trilinear $\gamma^\star$--D7-brane
couplings (\ref{d7couplings2},\ref{d7couplings4}).
On the other hand, this deformation can be forbidden by
certain discrete symmetries of the compact CY, but those symmetries would
only shift the mass spectrum of $(1/2,1/2,1)$ KK modes, not project them out
entirely.  If those KK modes are the lightest charged states (which may 
or may not be the case), the DM decays will be suppressed due to requiring
multiple insertions of other deformations (absent isometry-breaking D-branes, 
that is).

There are, of course, other isometry-breaking deformations, most of
which correspond to irrelevant operators in the gauge theory and
therefore decay exponentially in $z$.  If the lightest KK state
decays through a process  with a single deformation inserted, the
analysis will be very similar to that presented here.  However, due
to the difference in $z$ dependence and  coefficients, the
parametrics will differ from our results.  Some of these deformations
and decays, including amplitudes involving multiple  deformation
insertions, were discussed in \cite{aaron}.

Finally, the decays of KK modes in a throat that supports brane
inflation are important in the process of reheating.  These are the
same processes that we have discussed, but the throat is typically
much shorter,  $w\sim 10^{-3}$ or $10^{-4}$.

\section{Discussion}\label{s:discuss}

In this section we give a brief 
comparison to previous studies of KK modes in warped throats of string 
compactifications, then we outline some interesting future directions 
for the subject.
But first let us quickly recap our results.  Most of the decay 
channels we have considered, which should be representative of DM decays
into the given sectors, result in decay rates which are 
slow compared to the age of the universe.
Therefore, angularly charged KK modes are suitable dark matter candidates
in compactifications in which those decays dominate.  However, if the 
Standard Model is supported on D-branes with extent in the same warped 
throat that the KK modes occupy, the decay rates may be much faster.
For example, the natural decay rate onto a D3-brane that breaks the 
relevant angular isometries is $\tev$ scale, just set by the warped scale
at the bottom of the throat, which is also the mass scale of the KK mode.

On the other hand, having the SM supported on a D3-brane that
respects the  isometries of the DM candidate leads to an estimated
decay rate very similar to the observed bounds.  In particular, a
careful enough study may be able to constrain the parameter space of
D3-brane SM constructions in a class of compactifications.  A more
exciting possibility is that hints for decaying dark matter might
be confirmed in currently running experiments such as Fermi-LAT.   We can
make a similar statement for D7-brane SM  constructions that break
the appropriate isometries of the throat, though the relevant
parameter space has an extra dimension and the decay rates are more
sensitive to discrete compactification choices (represented by  $\nu$
and $\nu^\star$).

\subsection{Comparison to previous results}\label{s:compare}

As we discussed in the introduction, several groups have considered 
KK modes as DM candidates in string compactifications before. 
Proceeding thematically rather than chronologically, 
von Harling and Hebecker \cite{arXiv:0801.4015}
have considered KK modes of separate warped throats as DM candidates.
That is, they consider compactifications with multiple throats and take 
KK modes (without angular charge) as DM candidates.  They also considered
decays to the SM mediated by supergravity modes, which are similar in spirit
to the D7-brane SM decays mediated by the volume modulus in section 
\ref{s:d7decay}.  In addition, \cite{arXiv:0801.4015} use a gauge theory
dual to KS throats, which is appropriate in a different regime of validity than
the supergravity description of the throats.  In terms of the estimated
decay rates, however, the biggest difference with our results is that 
\cite{arXiv:0801.4015} does not account for the approximate isometry of the 
throats, which may lead them to overestimate some of the decay rates.
by overestimating the mixing with light supergravity modes.

Next, \cite{chentye} and \cite{dkp} have, between them, given a detailed 
cosmological history of KK modes, including charged KK modes as DM candidates,
but in a truncated theory with only KK modes of the 4D graviton.  Among 
decay channels, \cite{dkp} considered decays to a D3-brane SM sector; 
unsurprisingly, since the graviton couples to the brane scalar kinetic term
like our KK modes, they found similar decay rates to D3-brane degrees of 
freedom.  On the other hand, graviton KK mode interactions are somewhat
constrained by orthogonality relations, so we might expect correspondingly
richer physics in the thermal history of the full supergravity KK spectrum.

Our work is most closely related to that of \cite{aaron}.  That paper also
considered the full IIB supergravity and attempted to identify the lightest
charged state.  However, it did not account for contributions to the mass
beyond the geometrical contribution from $T^{1,1}$ dimensional reduction.  
Furthermore, it did not allow relevant (growing) deformations 
of the KS throat, which removed the $(1,0,0)$ $\Delta\Gamma$ deformation 
from consideration.  Therefore ref.\ \cite{aaron} did not consider as large a range
of decay channels as we have examined.  Lastly, \cite{aaron} was mostly concerned
with KK modes of a throat that supports brane inflation in order to 
determine whether charged KK modes yield dangerous relics which could
overclose the universe. A decay rate to gravitons was
found to be
\eq{aaron1}{\Gamma(\gamma^\star\to hh)
\approx w^{3.4}\frac{M_{3/2}^2k}{M_p^2}\ ,}
where $M_{3/2}$ is the amplitude of a supersymmetry-breaking deformation 
at the bulk ($z=0$).  Following \cite{hep-th/0411011}, generically the 
supersymmetry breaking scale satisfies $M_{3/2}^2 M_p^2\approx w^4 M_s^4$, 
so this decay rate becomes
\eq{aaron2}{\Gamma(\gamma^\star\to hh)\approx w^{7.4}\frac{M_s^4}{M_p^4}k
\ .}
This rate is matched or exceeded by a number of the decay channels we have 
considered, including all decays to SM modes on D3-branes, decays to
axions ($\gamma^\star\to a a^\star$), possibly gravitons
($\gamma^\star\to hh$, using our new estimate) and likely 
the direct decays to SM modes on D7-branes.  

Most recently, \cite{arXiv:0902.0008} constructed a model designed to
yield decay rates of the dark matter into SM particles at just the rate
needed to explain the astrophysical electron and positron excesses
seen by PAMELA, Fermi, ATIC, HESS, and other experiments
\cite{decays1,decays2}.  In that model, the dark matter candidate is
a KK mode of D7-brane fields  extending to the tip of a $\tev$ scale
throat.  The KK mode decays first to a light messenger, which then
decays to SM modes.  We have taken a different approach; rather than
design a model to give observable decay rates, we have instead
surveyed generic configurations in the class of conformally CY warped
compactifications of type IIB string theory for possible DM
candidates. We find that several of these backgrounds support DM
candidates with lifetimes of order $10^{25}$ s, as desired to match
the observed anomalies.

\subsection{Future directions}

We conclude by discussing several interesting directions for future work.
First, we have highlighted two sets of couplings of interest to string 
phenomenology.  One is the coupling of general supergravity fermions to 
brane fermions; see section \ref{sfkkm}.  The other, discussed in 
section \ref{s:d7spectrum}, is the coupling of D7-brane fermions to bosonic
supergravity fields in the presence of a general warp factor.  These 
couplings are necessary to complete our survey of dark matter candidates,
but they are also more generally of interest for the phenomenology of 
brane constructions.
In terms of decay rates for our KK DM candidates, it would also be 
worthwhile
to determine the effects of approximate isometries and their breaking in
the bulk CY on tunneling rates.  

The obvious next step is to detail the cosmic history of the dark matter
candidates we have discussed in this paper, as was done for graviton KK 
modes in refs.\ \cite{chentye,dkp}.  This would require estimating the couplings
between the various charged and uncharged KK modes and the moduli.  While
we were able to ignore many of the fields of type IIB supergravity in
calculating decay rates, thermalization rates are sensitive to interactions
among all the fields.  This is therefore a more ambitious project.  
In addition, it may be necessary to understand the full dynamics of long
throats during high-scale inflation and reheating, as emphasized in
refs.\ \cite{hep-th/0501184,fmm}.  Some steps have been taken in this
direction by refs.\ \cite{gm,stud,ftud}, although it seems likely that the 
effects of the inflationary potential will need to be incorporated in the 10D
theory, rather than just the 4D effective theory.

Such a careful accounting of cosmological history might provide
constraints on the parameter space of D-brane Standard Model 
constructions with warped hierarchies.  Assuming the charged KK modes
are  populated with the appropriate density, there are both D3-brane
and D7-brane models that yield observable (or nearly observable)
decay rates in  natural regions of parameter space.  
It is rare to have the opportunity to constrain compactifications of
string theory, based on considerations which are independent of the need
for embedding the SM; this would give an additional, independent criterion   
for ruling out regions of the string landscape.
Therefore, it would be valuable to study
both isometry-preserving D3-branes and isometry-breaking D7-branes to
see how strongly their parameter spaces are constrained.

\acknowledgments
We would like to thank Mariana Gra\~na and Bret Underwood for useful 
discussions.  As parts of this work have been presented previously, we 
would also like to thank the various audience members for their questions
and comments (at UMass Amherst, MIT/Tufts/CfA cosmology seminar, Carleton
University, and the Perimeter Institute Holographic Cosmology conference).
Our work has been supported by NSERC.

\appendix

\section{D7-brane induced metric for Kuperstein embedding}\label{s:d7metric}

In this appendix, we demonstrate that the D7-brane scalar kinetic
term (\ref{d7kinetic}) is a good approximation for a D7-brane that obeys 
the Kuperstein embedding \cite{kup}.  
Such D7-branes are supersymmetric and also
have been shown to generate inflationary potentials for probe D3-branes
\cite{inflection1,inflection2}.  Recently, \cite{d7susy} studied the 
excitations of Kuperstein-embedded D7-branes and their couplings to closed
string modes; we follow a simplified version of their results.

The metric of the conifold was given as a foliation
of Kuperstein embeddings in \cite{d7susy}; this is a convenient set of 
coordinates to use since we consider fluctuations around a fixed Kuperstein
embedding.  Therefore, in D7-brane calculations, rather than (\ref{gtilde}),
we use 
\bea
d\t s^2 &=& \frac{k}{3}e^{kz} \left\{ \frac{|\mu+\chi|^2}{2}\left[
\frac{1+2\cosh\rho}{3}(d\rho^2 +h_3^2) +\cosh^2\left(\frac{\rho}{2}\right)
h_1^2 +\sinh^2\left(\frac{\rho}{2}\right)h_2^2\right]\right.\nonumber\\
&&\left. +\frac{2}{3}\sinh\rho (d\rho+ih_3)(\mu+\chi)d\b\chi +\textnormal{c.c.}
+\frac{4}{3}(1+\cosh\rho)d\chi d\b\chi\right\}\ .
\label{conkup2}\eea
(See equation (C.8) of \cite{d7susy} with the supersymmetry breaking parameter
$\mathcal{S}=0$.)
Our usual radial coordinate $z$ is related to these coordinates by
\eq{radialkup}{\frac{e^{-3kz}}{k^3} = |\mu+\chi|^2 (1+\cosh\rho)\ .}
In the above, the $h_i$ are angular differentials for the angular directions
along the brane, and 
$\mu$ is a constant that chooses the location of the embedded
D7-brane.  We can expand around $\chi=0$, since changing $\chi$ just shifts
$\mu$ ($\mu$ acts as the expectation value of $\chi$).  In fact, since we
are not interested in self-interactions of $\chi$, we can set $\chi=0$
while retaining $d\chi$, and we do so in the following.

A non-fluctuating brane ($\chi=0$) attains its maximum value of $z=z_1$ 
at $\rho=0$, and we find that
\eq{muw1}{\mu^2 = \frac{w_1^3}{2k^3}\ ,\ \ w_1\equiv e^{-kz_1}\ .}
Furthermore, just as we approximate the tip of the KS throat with a 
boundary condition at $z_0$ for the $\ads\times T^{1,1}$ metric, which is
valid at $z\ll z_0$, we approximate (\ref{conkup2}) with a metric
valid at $z\ll z_1$ and replace the smooth metric near $z_1$ with a radial
boundary condition.  Since small $z$ is large $\rho$, we can replace
$\cosh\rho$ and $\sinh\rho$ with $e^\rho/2$ and further $d\rho\sim -3kdz$. 
Then the $d\chi$ terms in (\ref{conkup2}) become 
\eq{conkup3}{d\t s^2 \approx 
\frac{2}{9}\frac{\sqrt{2k}}{w_1^{3/2}} e^{-2kz}\left(-3dz+i\frac{h_3}{k}\right)
d\b\chi +\textnormal{c.c.}
+\frac{8}{9}\frac{k}{w_1^3} e^{-2kz}d\chi d\b\chi+\cdots\ ,}
where we have placed a factor of $1/k$ with the angular differential.

Finally, we need to rescale $\chi,\b\chi$ to match our usual conventions.
First, $\chi$ as written has dimensions of $\textnormal{length}^{3/2}$. 
To get dimensions of length, we rescale to $\chi'=\sqrt{k}\chi$.  The 
other rescaling we must do is more subtle.  The point is that, as $z\to 0$,
$\t g_{mn}$ given in (\ref{conkup3}) should match smoothly to a CY metric
$\t g_{mn}$ with components generically order 1.  However, all terms of 
(\ref{conkup3}) with a factor of $d\chi$ or $d\b\chi$ are larger than order
one by a factor of $w_1^{-3/2}$ for each factor of $d\chi$.  Therefore,
the correct coordinate for matching onto the bulk CY metric is
$\chi''=\chi'/w_1^{3/2}$. (Note that the $\rho$ and angular components of 
(\ref{conkup2}) are order 1 as $z\to 0$, so those coordinates do not
need to be rescaled.)

In the end, we find
\eq{conkup4}{d\t s^2 \approx 
\frac{2\sqrt{2}}{9}e^{-2kz}\left(-3dz+i\frac{h_3}{k}\right)
d\b\chi'' +\textnormal{c.c.}+\frac{8}{9}e^{-2kz}d\chi'' d\b\chi''+\cdots\ .}
In the full 10D metric, this is multiplied by the warp factor 
$e^{-2A}=e^{2kz}$.  Therefore, the pull-back metric to the D7-brane in
static gauge is
\eq{conkup}{ds^2=\left[e^{-2kz}\eta_{\mu\nu}+\frac{8}{9}
\del_{(\mu}\chi\del_{\nu)}\b\chi\right]dx^\mu dx^\nu+\left[
\frac{2\sqrt{2}}{9}\left(-3dz+i\frac{h_3}{k}\right)
\del_\mu\b\chi dx^\mu+\textnormal{c.c.}\right]
+\cdots\ ,}
where the $\cdots$ represent terms containing only $dz$ and $h_i$ (and
which have no powers of $e^{-kz}$).
The kinetic term, as usual, comes from expanding the determinant of the metric
(\ref{conkup}) to first order in $|\del\chi|^2$.
These terms come from the trace of the fluctuations in the first square
brackets of (\ref{conkup}) (after factoring out the warped Minkowski metric)
and also from the square of the off-diagonal terms in the second square
brackets (contracted with the warped Minkowski metric).  Both of these terms
carry a factor of $e^{2kz}$ and are both multiplied by the determinant of the
unfluctuated 8D metric, which carries a factor of $e^{-4kz}$.  This yields
an overall kinetic term $e^{-2kz} |\del\chi|^2$ to be integrated over the
compact dimensions of the D7-brane, in agreement with (\ref{d7kinetic}) in
the throat.

\bibliographystyle{utcaps2}

\bibliography{kkdm}

\end{document}